\documentclass[%
 reprint,
superscriptaddress,
 amsmath,amssymb,
 aps,
 prd,
floatfix,
]{revtex4-2}

\usepackage{graphicx}%
\usepackage{dcolumn}%
\usepackage{bm}%
\usepackage{physics}%
\usepackage{mathtools}%
\usepackage{hyperref}%
\usepackage{ulem}

\newcommand{\chieff}{\chi_\text{eff}}
\newcommand{\Xeff}{\chi_\text{eff}}
\newcommand{\Mo}{M_\odot}

\usepackage{color}
\usepackage{xcolor}

\newcommand{\refine}[1]{{{#1}}}

\newcommand{\figref}[1]{Fig. \ref{#1}}
\newcommand{\Belcpop}{Pop.~I+II~}

\usepackage{acronym}
\newacro{GW}[GW]{Gravitational Wave}
\newacro{BBH}[BBH]{Binary black hole}
\newacro{PopI}[Pop.~I]{Population I}
\newacro{PopII}[Pop.~II]{Population II}
\newacro{PopIII}[Pop.~III]{Population III}
\newacro{GC}[GC]{Globular cluster}
\newacro{NSC}[NSC]{Nuclear star cluster}
\newacro{CHE}[CHE]{chemically homogeneous evolution}
\begin{document}

\preprint{APS/123-QED}

\title{Constraint on the progenitor of binary black hole merger using Population~III star formation channel}

\author{Masaki Iwaya}
 \email{iwayam@cardiff.ac.uk}
\affiliation{
 Institute for Cosmic Ray Research, University of Tokyo, Kashiwanoha 5-1-5, Kashiwa, Chiba 277-8582 Japan
}
\affiliation{
School of Physics and Astronomy, Cardiff University, Cardiff, CF24 3AA, UK
}

\author{Tomoya Kinugawa}
\affiliation{Faculty of Engineering, Shinshu University, 4-17-1, Wakasato, Nagano-shi, Nagano, 380-8553, Japan}
\affiliation{Research Center for Advanced Air-mobility Systems, Shinshu University, 4-17-1, Wakasato, Nagano-shi, Nagano, 380-8553, Japan}
\affiliation{Research Center for the Early Universe, Graduate School of Science, University of Tokyo, 7-3-1 Hongo, Bunkyo-ku, Tokyo 113-0033, Japan}

\author{Hideyuki Tagoshi}
\affiliation{
 Institute for Cosmic Ray Research, University of Tokyo, Kashiwanoha 5-1-5, Kashiwa, Chiba 277-8582 Japan
}

\date{\today}%

\begin{abstract}
The observations of gravitational waves have revealed the existence of black holes (BHs) above $30\Mo$. A variety of channels have been proposed as their origin, including the \ac{PopIII} star channel. In this channel, \ac{BBH} containing such massive BHs are naturally produced. In this paper, we examine the relative fractions of five formation channels that may contribute to the origins of \ac{BBH}s: isolated binaries of either Population~I or Population~II stars, \ac{PopIII} isolated binaries, chemically homogeneous evolution, and the dynamical evolution in globular clusters and nuclear star clusters, using the LIGO-Virgo-KAGRA gravitational-wave transient catalog (GWTC-3) events through hierarchical Bayesian inference. We find that the branching fraction of the \ac{PopIII} \ac{BBH} channel is $0.11^{+0.08}_{-0.06}$ within our framework, consistent with the local merger rate density of the model of the \ac{PopIII} \ac{BBH} channel we adopt. We also evaluate the contributions to the catalogue using the selection effect of each formation channel and find that \ac{PopIII} \ac{BBH} could contribute at a non-negligible rate, though the adequacy of these ratios should be subject to ongoing discussion.

\end{abstract}
\maketitle

\section{Introduction}
Since the first direct detection of the \ac{GW} \cite{GW150914}, the LIGO-Virgo-KAGRA (LVK) network has observed $\sim$300 candidates of compact binary coalescences by their third observing run \cite{GWTC-1, GWTC-2.1, GWTC-3, GWTC-4}, most of them considered to be mergers of \ac{BBH}. These observations provide a brand-new approach to inspecting the origin of stellar-mass black holes. 

The masses of black holes of the first gravitational wave event, GW150914, were about $\sim30\Mo$, and the existence of such BHs was a surprise since there were few predictions of such a massive BH by that time. This observation sparked a discussion about the origin of \ac{BBH}.

There are two main channels proposed for the formation of merging \ac{BBH}s: the isolated formation channel and the dynamical formation channel.
In the isolated formation channel, the progenitors of binary systems are gravitationally bound from the zero-age main-sequence phase,  and they evolve without external disturbances and eventually become merging compact objects. Binary interactions are vital for coalescence within the Hubble time. The channel is finely divided into sub-channels, depending on how the binaries interact. In the classical picture of the channel, a binary undergoes an unstable mass transfer and a common envelope phase, in which the binary drastically loses its orbital energy (for example, \cite{CE1, CE2, CE3}). Recently, however, it was reported that the merging \ac{BBH}s can also be produced through stable mass transfer (for example, \cite{SMT1, SMT2}).

The stellar metallicity is one of the elements that affects the stellar evolutionary tracks, which in turn affects the BHs that remain as stellar remnants.
The possibility that the isolated binaries of \ac{PopIII} stars are the origins of the \ac{BBH}s has been a subject of much discussion (for example, \cite{K2014,popIII-1}).
\ac{PopIII} stars are the first stars in the universe, with zero metal (or extremely low-metal). The evolution of these stars differs from that of \ac{PopI} stars or \ac{PopII} stars.
\refine{\ac{PopI} stars are relatively metal-rich stars in the Galactic disk, typically with metallicities ranging from approximately 1/10th to three times that of the Sun \cite[e.g.][]{1944ApJ...100..137B,2015ARA&A..53..631F}, while \ac{PopII} stars are the low metal stars with 0<Z$\lesssim 0.1Z_\odot$, where $Z_\odot$ denotes the solar metallicity. The threshold separating \ac{PopII} and \ac{PopIII} was reported as $ Z\sim10^{-5}Z_\odot$ \cite{Tanikawathreshold}.}
As the progenitor of \ac{BBH}, BHs from \ac{PopIII} stars tend to have higher masses \refine{than those from \ac{PopI}/\ac{PopII} stars on average}, typically $\sim30\Mo$. There are reasons for this tendency: \refine{first, the initial mass function of \ac{PopIII} stars is expected to be more top-heavy than that of \ac{PopI} and \ac{PopII} stars, due to the inefficient cooling of the primordial gas \cite{Bromm2004}.} \refine{Also, binaries of \ac{PopIII} tend  not to lose stellar mass through binary interactions because they tend to evolve via blue giant stars with a small radius \cite{K2014,K2016,Inayoshi2017,K2020chirp}. Moreover, the mass loss by stellar winds is greatly suppressed for no metal stars \cite{Marigo2003}.}

%On the other hand, in the dynamical formation channel, merging \ac{BBH}s are formed in the dense region, where interaction with a binary system with other members of the region is possible. Since different types of dense regions are expected to have different parameter distributions of \ac{BBH}s, numerous investigations have been performed for dense stellar environments \cite{GC1, GC2, YSC1}.

\refine{On the other hand, several models consider dynamical formation
in dense stellar environments.
Binary black holes can form efficiently in globular clusters
\cite[e.g.][]{GC1,GC2,Rodriguez2016},
in young stellar clusters
\cite[e.g.][]{Banerjee2017, DiCarlo2019,YSC1},
and in nuclear star clusters
\cite[e.g.][]{Antonini2016, Antonini2019}.
In these environments, repeated dynamical interactions
can assemble and modify \ac{BBH}s prior to merger.}

\refine{Other formation channels, such as the \ac{CHE}  \cite{CHE1,Marchant2016,CHEmodel}, dynamical triples \cite{Triple1,Silsbee2017}, the formation in active galactic nuclei \cite{AGN1, AGN2,Stone2017,Tagawa2020}, and primordial black holes (PBHs) \cite{Sasaki2016,PBH1,2017PhRvD..96l3523A} have also been proposed.
The merger rate predictions of \ac{BBH}s for various formation channels are summarized in \cite{Mandel2022review}.}

The \ac{BBH} formation channels must be consistent with observational results. The merger rate of \ac{BBH}s, and the distribution of parameters of \ac{BBH}s inferred from the observation, are key elements to constrain the \ac{BBH} formation channels. 
LVK found the overdensities in the mass distribution near $\sim 10\Mo$ and $\sim 35\Mo$, \refine{the population-level anti-correlation between mass ratio and effective spin parameter which was pointed out first in \cite{qXeffCorrelation}}, and the redshift evolution of the merger rate density. \refine{Those properties need to be explained by formation channels.}

%There have been attempts to constrain the formation channels of the GW sources based on the GW catalogue.
%\textcolor{red}{TODO: Add works that attempt to clarify the origins with concrete scenarios.}
Identifying the formation channels of the \ac{GW} sources is one of the central goals of \ac{GW} astronomy. A number of studies have attempted to constrain these channels using the growing \ac{GW} catalogue.
On the theoretically-driven approach, population-synthesis models and/or dynamical formation scenarios have been compared against the observed distributions of source parameters. For instance, Ref. \cite{Zevin2021} combined five state-of-the-art \ac{BBH} formation pathways and concluded that multiple formation pathways and proper physical prescriptions are needed to interpret the catalogue. 
In Ref. \cite{Franciolini2022}, the PBH channel is explored with a focus on the correlation in the mass-ratio parameter
\begin{equation}
q = \frac{m_2}{m_1} (\leq1),
\end{equation}
and the effective inspiral spin parameter of the \ac{BBH}
\begin{equation}
    \chieff = \dfrac{(\bm{\chi}_1+q\bm{\chi}_2)\cdot\hat{\bm{L}}}{1+q},
\end{equation}
where $m_1$ and $m_2$ are the component masses of the \ac{BBH} with $m_1\geq m_2$, $\bm{\chi}_1$ and $\bm{\chi}_2$ are the dimensionless spin parameters of each BH, and $\hat{\bm{L}}$ is the unit vector parallel to the orbital angular momentum.
In parallel, data-driven approaches that rely on minimal astrophysical assumptions on the source-parameter distribution shapes, such as \cite{NonparaCallister}.
The examples cited above represent only a fraction of the rapidly expanding literature on this topic. For a broader review, see, e.g., \cite{MapelliReivew, MandelFarmerReview}.

In the \ac{GW} astronomy context, \ac{PopIII} stars have attracted interest since the first direct \ac{GW} observation. The mass range of the first event, GW150914, is both $\sim 30 \Mo$ \cite{GW150914}.
However, the absence of direct observation of these stars makes the theory uncertain. This uncertainty results in uncertainty in the star formation rate of these stars. Estimated local merger density of \ac{BBH}s from \ac{PopIII} stars spans from $10^{-2}\ \mathrm{Gpc^{-3}yr^{-1}}$ \cite{Hartwig} to $10^{2}\ \mathrm{Gpc^{-3}yr^{-1}}$ \cite{K2020chirp}. 
In addition, since the \ac{PopIII} stars are the first stars in the universe, the merger rate of \ac{PopIII} \ac{BBH}s starts to reach a maximum at a high redshift, for which only next-generation gravitational wave detectors can observe. Therefore, there are proposals \cite{K2014, PopIIIBelc, K2020chirp,2023popIII} to verify the channel by using the next-generation detectors such as Einstein Telescope \cite{ET}.

\refine{Although there is large uncertainty in the prediction of the merger rate of the \ac{PopIII} \ac{BBH}s in various computations
\cite{K2014,2023popIII}, binary population synthesis studies have shown that the BH mass distribution from \ac{PopIII} binaries
typically exhibits a peak around $\sim 30\Mo$. In addition, more massive \ac{BBH} mergers, such as GW190521, can also be interpreted as originating from \ac{PopIII} \ac{BBH}s \cite[e.g.][]{GW190521popIII,Farrell:2020zju}.}
\refine{\ac{PopIII} stars are difficult to observe directly with optical and infrared telescopes, as they formed and likely ended their lives at very high redshift.} Thus, it is important to constrain their properties and the formation history with \ac{GW} observations.

In this work, we consider a binary population model consisting of a mixture of five \ac{BBH} population models: isolated binaries from either \ac{PopI} or \ac{PopII}, isolated binaries from \ac{PopIII}, and \ac{BBH}s dynamical evolution either in \ac{GC} or \ac{NSC}, and \ac{CHE}. We perform a hierarchical Bayesian analysis 
by using the \ac{BBH} events listed in the GWTC-3 catalog by LVK. We then constrain the fraction of the contribution
of each formation channel to the cosmic merger rate and the observed merger rate of \ac{BBH}s. 
We also investigate the distribution of parameters that describe the \ac{BBH} waveform. 

The paper is organized as follows. Sec.\ref{Sec.Met} describes the formation models we adopt in this work and the summary of the hierarchical Bayesian inference. In Sec.\ref{Sec.Res}, we describe the result of our inference, namely the posterior distribution for branching fraction and the Bayes factors, in Sec.\ref{Sec.Dis}, we discuss further the predicted parameter distribution and the constraint on the merger rates. In Sec.\ref{Sec.Sum}, we summarize our findings.

\section{Methods}
\label{Sec.Met}

In this section, we summarize the formation channel models we adopt and the hierarchical Bayesian analysis framework.

\subsection{BBH channels}

This study utilizes sample sets of \ac{BBH} systems obtained from population syntheses from several published formation models:

\begin{itemize}
\item Isolated evolution of \ac{PopI} and \ac{PopII} stars
\item Isolated evolution of \ac{PopIII} stars
\item Chemically homogeneous evolution
\item Dynamical evolution in the globular cluster
\item Dynamical evolution in Nuclear stellar cluster
\end{itemize}

For the isolated evolution of \ac{PopI} and \ac{PopII} stars (hereafter \Belcpop channel), we use the result from \citep{Belczynskimodel}. Among their models, we use the model called M30.B, which is the standard model in their work. This model employed state-of-the-art physical models such as star formation rates \citep{SFR} and an accretion model. The model assumes the efficient angular momentum transport that includes the Tayler-Spruit magnetic dynamo \citep{Spruit} so that the natal spin magnitude for BH is in the range $0.05\lesssim a\lesssim 0.15$.  The \ac{BBH} local merger rate of the model is 43.7 $\mathrm{Gpc^{-3} yr^{-1}}$. 

For the isolated evolution of \ac{PopIII} stars (hereafter \ac{PopIII} channel), we use a model presented in \citep{K2020chirp}.
The model we adopt is the `M100' model in \citep{K2020chirp}, which was shown to fit well with the GWTC-2 results in \citep{kinugawa2021}. The model assumes a flat initial mass function from $10\Mo$ to $100\Mo$ for Pop~III binaries. The local merger rate in this model is 6.36 $\mathrm{Gpc^{-3} yr^{-1}}$. 

For the remaining formation channels, the results of population syntheses are taken from the previous work \citep{Zevin2021}. More particularly, \citep{CHEmodel} for \ac{CHE}, \citep{R'sGC} for dynamical evolution in \ac{GC}, and \citep{Antonini2019} for dynamical evolution in \ac{NSC}. All the model assumes the natal spin of the BH is zero. We will denote them as \ac{CHE}, \ac{GC}, \ac{NSC} channel resepectively.

\subsection{Population Inference}
In this work, we assume that \ac{BBH}s in the universe are formed in several different formation channels. We then consider a mixed model in which the total merger rate is the sum of the merger rates of each channel weighted by the branching fractions $\{f_j\}$  with $\sum_{j} f_j = 1$, where $j$ denotes each formation channel. 
Here, $f_j$ represents the underlying contribution of the $j$-th channel to the astrophysical BBH population, corrected for observational selection effects.
We apply a hierarchical Bayesian method \cite{HBcite, HBcite2} to estimate the branching fractions 
by using the \ac{BBH} events in the GWTC-3 catalog \cite{GWTC-3}.
We consider \ac{BBH}s in GWTC-3 with the False Alarm Rate less than 1 $\mathrm{yr^{-1}}$. This gives us 69 events \cite{GWTC-3:pop}. We use the posterior/prior samples that are available at the Gravitational Wave Open Science Center (GWOSC) \cite{GWOSC}.

In this work, we don't estimate the total merger rate itself. Therefore, we marginalize the total coalescence rate of \ac{BBH}s by assuming a log-flat prior \cite{HBcite}.
In that case, we can deduce the posterior probability density of branching fraction $\Lambda = \qty{f_j}$, $p(\Lambda\mid\qty{x}),$ given observation set $\qty{x}$:
\begin{equation}
\label{eq:final}
p(\Lambda\mid\qty{x})\propto\pi(\Lambda)\prod_{i=1}^{N_\mathrm{obs}}\dfrac{1}{\alpha(\Lambda)}\int\dd{\theta_i}\dfrac{p(\theta_i\mid x_i)}{\pi(\theta_i)}p(\theta_i\mid\Lambda),
\end{equation}
The posterior probability density describes the probability density given the observations $\qty{x}$ to infer the unknown population parameter $\Lambda$ through the binary parameters $\theta$.
A detailed derivation of this equation is given in the Appendix \ref{Sec:HBDerivation}.
The $\alpha(\Lambda)$ in this equation is a quantity that can be called "detection efficiency" and will be explained later. $p(\theta_i\mid x_i)$ is the posterior probability density for estimating the source parameters based on the observed data $x_i$, and $\pi(\theta_i)$ 
is the prior probability density of the parameter $\theta_i$.
$p(\theta_i\mid\Lambda)$ is the probability density that describes the probability of an event occurring when the hyperparameter is $\Lambda$ and the true event parameter is $\theta$. Since the integral on the multidimensional parameter space is computationally expensive, 
we approximate the integral by the discrete sum of the posterior samples of each event available in the LVK GWTC-3 at GWOSC. We have
\begin{equation}
\label{eq:approxed}
p(\Lambda\mid\qty{x})\propto\pi(\Lambda)\prod_{i=1}^{N_\mathrm{obs}}\dfrac{1}{\alpha(\Lambda)}\dfrac{1}{S_i}\sum_{k=1}^{S_i}\dfrac{p(\theta_i^k\mid\Lambda)}{\pi(\theta_i^k)},
\end{equation}
where $S_i$ denotes the total number of posterior samples used in calculating the sum, and $\qty{\theta_i^k}$ are the LVK posterior samples.

The last term $p(\theta_i\mid\Lambda)$ in (\ref{eq:final}) is given as 
\begin{equation}
p(\theta\mid\Lambda) = \sum_j f_j p(\theta\mid \mu^j),
\end{equation}
where $p(\theta\mid \mu^j)$ is a probability density distribution that describes the probability that a source with the parameter $\theta$ is generated in $j$-th formation channel, represented by $\mu^j$.

Under this formula, we can express $p(\Lambda\mid\qty{x})$ as
\begin{equation}
p(\Lambda\mid\qty{x})\propto\prod_{i=1}^{N_\mathrm{obs}}\dfrac{1}{\alpha(\Lambda)}\sum_j\dfrac{f_j}{S_i}\sum_{k=1}^{S_i}\dfrac{p(\theta_i^k\mid\mu^j)}{\pi(\theta_i^k)}.
\end{equation}

In this work, we use four parameters for $\theta$: the source frame chirp mass $\mathcal{M},$ the mass ratio $q,$ the effective inspiral spin parameter $\chi_\mathrm{eff}$, and the merger redshift $z$.
 
The detection efficiency $\alpha(\Lambda)$ is computed as
\begin{align}
\alpha(\Lambda) &= \int\dd{\theta} p(\theta\mid\Lambda)p_\mathrm{det}(\theta) \notag\\&= \sum_{j}f_j\int\dd{\theta} p(\theta\mid\mu^j)p_\mathrm{det}(\theta) \eqqcolon \sum_j f_j\alpha_j,
\end{align}
where we set
\begin{equation}
\alpha_j \coloneqq \int\dd{\theta} p(\theta\mid\mu^j)p_{\mathrm{det}}(\theta).
\end{equation}
In this equation, $p_\mathrm{det}(\theta)$ is the probability that a binary merger with a true event parameter of $\theta$ can be detected in noisy observations:
\begin{equation}
p_\mathrm{det}(\theta) = \int\dd{x} p(x\mid\theta),
\end{equation}
where the integration sums up all detectable \ac{GW} signals.

In this work, we use the injection set from GWTC-3 \cite{GWOSC} to evaluate the detection efficiency by using the importance sampling method. We also tested the evaluation method applied in \cite{Zevin2021} and found that the evaluated values are consistent with our method.

Recent population studies have some criteria for $\alpha(\Lambda)$ to reject unphysical hyperparameters (e.g., \cite{Precision_Requirement}), but we do not have them because imposing those restrictions would severely limit the possible Hyperparameters.
For $p(\theta\mid\mu^j)$, we construct the kernel density estimator (KDE) from the calculations of the $j$-th astrophysical channels. Some of the formation channels have a peak at $q=1$, which means that several formation channels tend to have equal-mass binary systems. However, since $q\leq 1$ is a physical constraint of the parameter, the simple KDE method fails to represent the distribution around $q=1$. Therefore, we applied the reflection method \cite{Silverman} to the data point $q=1$. Finally, we calculate the prior distribution $\pi(\theta)$, which was used by LVK for each \ac{BBH} candidate. For the event prior to GWTC-3, $\pi(\theta)$ is uniform in redshifted component mass, uniform in spin magnitude, and isotropic in spin orientation. There are two variants of posterior samples for each event, the difference being the distance prior. As in \cite{GWTC-3:pop}, we use the `nocosmo' file to rely on the abundant number of posterior samples \cite{GWTC-1, GWTC-2.1, GWTC-3}.
%Now, we can estimate the distribution of $p(\Lambda\mid\qty{x})$.
We utilize the \texttt{Dynesty} package for plotting the posterior distribution. 
%The Bayes factor between the models $\mathcal{M}_1$ and $ \mathcal{M}_2$ is given as
%\begin{equation}\mathcal{B}_{\mathcal{M}_1}^{\mathcal{M}_2} = \dfrac{\int\dd{\Lambda}p(\qty{x}\mid\Lambda)\pi(\Lambda\mid\mathcal{M}_2)}{\int\dd{\Lambda}p(\qty{x}\mid\Lambda)\pi(\Lambda\mid\mathcal{M}_1)}.\end{equation}

\section{Results}
\label{Sec.Res}
\begin{figure*}[htb!]
\begin{center}
\includegraphics[width=15cm]{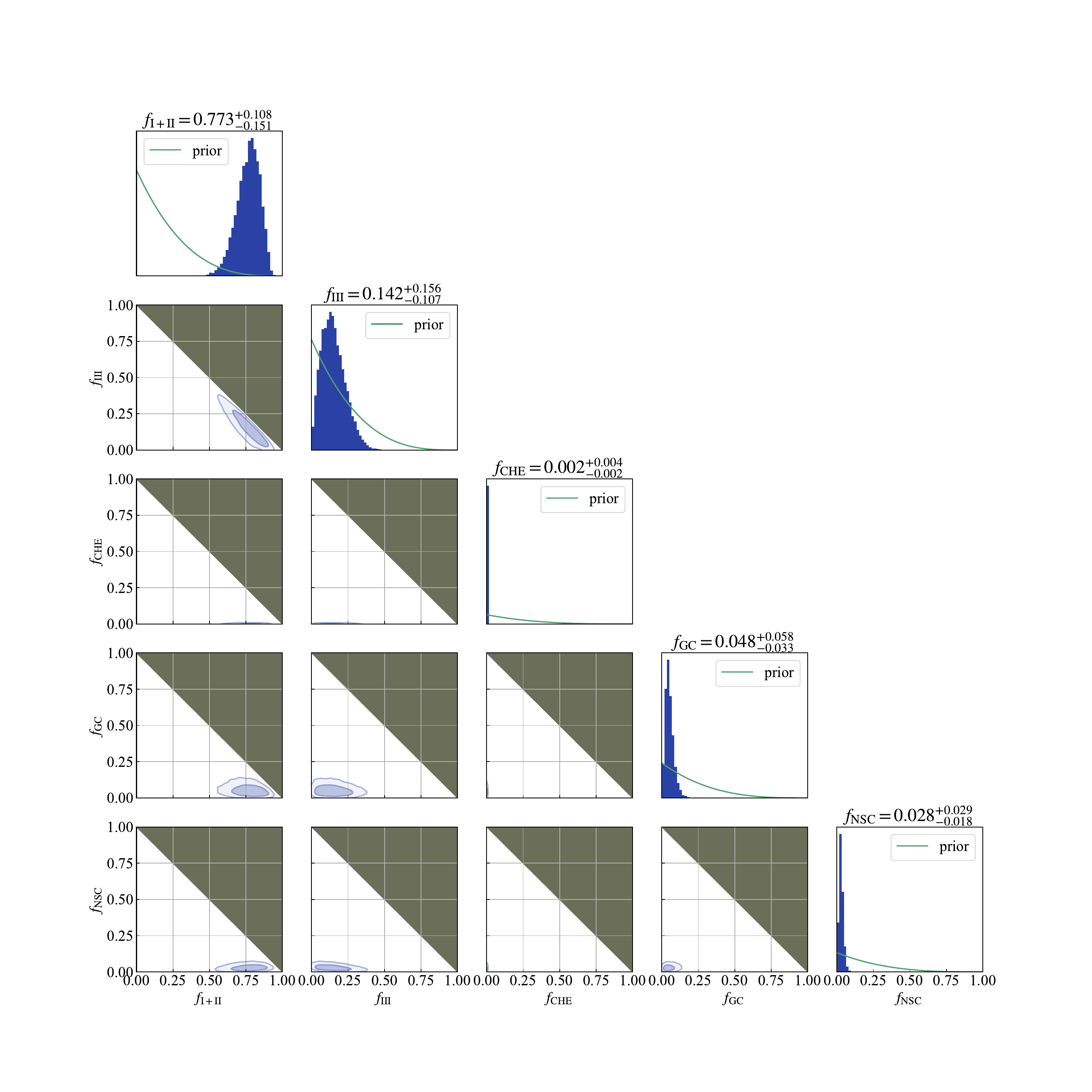}
\caption{The hyper-posterior distributions for underlying fractions $\{f_j\}$. The dark regions are excluded by the simplex constraint $\sum_j f_j = 1,\ f_j \geq 0$. In the two-dimensional plot, the $1\sigma$ and $2 \sigma$ regions are shown by contours.}
\label{fig: Fractions}
\end{center}
\end{figure*}
The result from a synthesis of five formation channels is shown in \figref{fig: Fractions}. 
\refine{This shows the inferred underlying fraction, representing the intrinsic contribution of each formation channel to the astrophysical BBH population in the Universe.}
In this mixing scheme, the dominant formation channel is \Belcpop, with an underlying fraction of $0.77^{+0.11}_{-0.15}$ at the 90\% confidence interval. On the other hand, if one takes the \ac{PopIII} channel into account, the branching fraction of this formation channel accounts for $0.14^{+0.16}_{-0.11}$ of the total. \ac{CHE} channel contributes quite a low percentage in this synthetic scheme, with $<0.6\%$ contribution estimated at the 90\% confidence interval.
The contribution from dynamical evolutions, evolution in GC or NSC, accounts for $0.08^{+0.07}_{-0.04}$, and the contribution from the GC is greater than that of the NSC.

\begin{figure*}[htb!]
\begin{center}
\includegraphics[width=15cm]{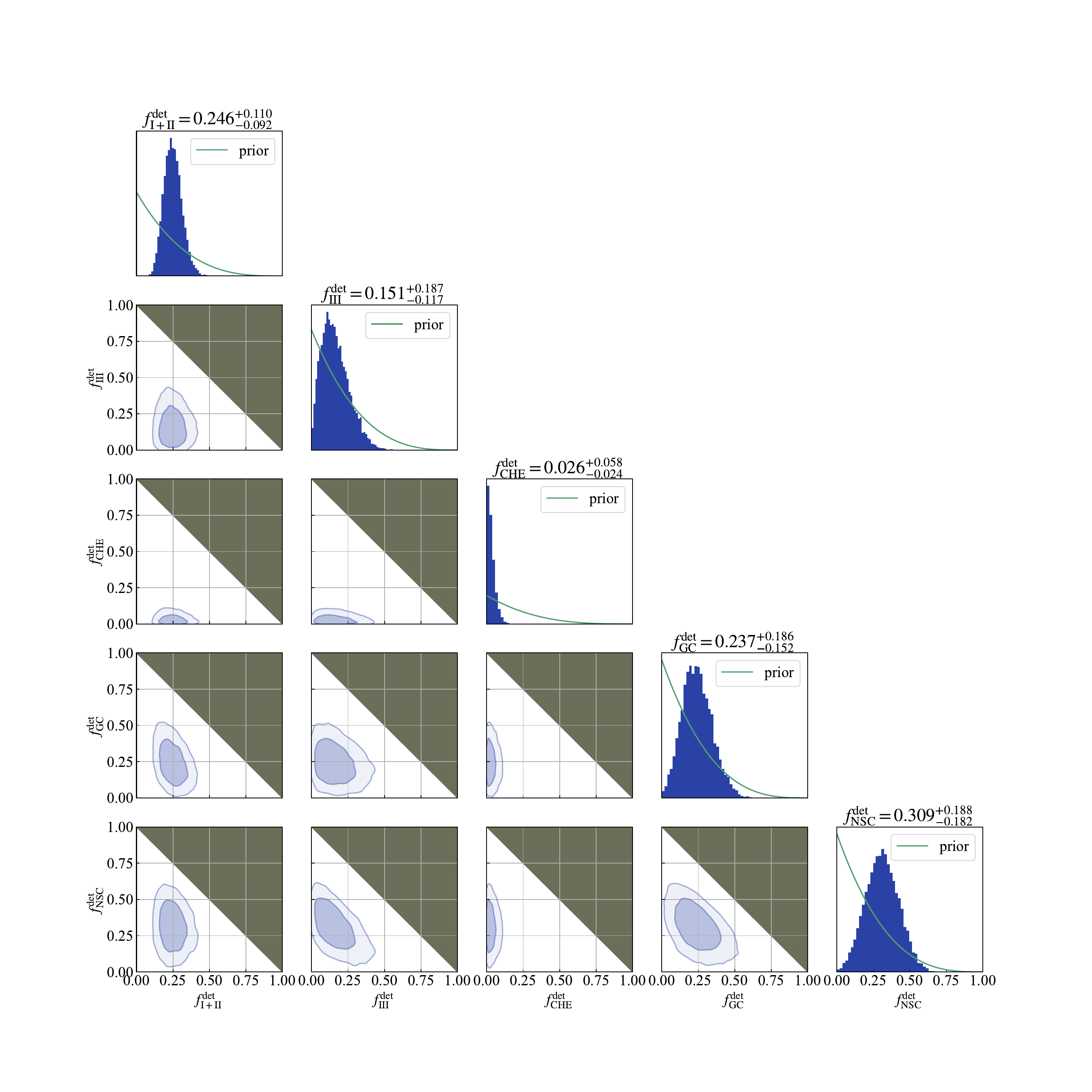}
\caption{Same as \ref{fig: Fractions}, but for detectable fraction defined as \eqref{eq: detectable_fraction}; In other words, these fraction parameters indicate the contribution from each channel for LVK-observable population.}
\label{fig: Observable_Fractions}
\end{center}
\end{figure*}

From the sensitivity estimation of the model and the underlying fraction parameter, one can define the detectable fraction of a channel \citep{Zevin2021}:
\begin{equation}
\label{eq: detectable_fraction}
f^\mathrm{det}_j = \frac{f_j\alpha(\mu_j)}{\sum_kf_k\alpha(\mu_k)},
\end{equation}
which indicates the contribution to the observed \ac{GW} catalog from each population channel. \figref{fig: Observable_Fractions} demonstrates the 5-dimensional distribution in terms of detectable fraction. We see that the contribution from \Belcpop channel is disfavored in the detectable fractions. This is because most of the \ac{BBH}s that can be made in this channel are less massive, so the selection effect is relatively small. The selection effect in this calculation for this channel is $\alpha(\mu_\mathrm{I+II}) = 3.4\times10^{-4}$, which is at least 10 times smaller than other formation channels. Instead, the contribution from dynamical formation channels is pushed in terms of detectable fractions. The contribution from the \ac{CHE} channel to the detectable fraction is zero-consistent, but it rises to $0.03^{+0.06}_{-0.02}$ at 90\% credibility. 

To evaluate the presence of the \ac{PopIII} formation channel, we compared two nested models: A model that incorporates five channels, and a model in which the \ac{PopIII} formation channel does not contribute to the catalog at all. We evaluate the likelihood ratio among these models and find that the existence of \ac{PopIII} star formation channel improves the likelihood by $11.87,$ or $1.07$ in $\log_{10}$ scale. This big difference suggests that the \ac{PopIII} formation channel captures important aspects of GWTC-3 that cannot be adequately represented by other channels.
 
In this alternative scheme, the domination of the isolated evolution of \Belcpop is much stronger, with a branching ratio of $0.89^{+0.05}_{-0.08}$ at the 90\% confidence interval, while the contribution from dynamical evolutions is also greater, $0.07^{+0.07}_{-0.04}$ and $0.04^{+0.03}_{-0.02}$ for GC and NSC, respectively.

\section{Discussion}
\label{Sec.Dis}

\refine{We analyze the latest \ac{BBH} merger catalog to infer the fractions of astrophysical \ac{BBH} formation channels, including the isolated evolution of \ac{PopIII} stars.
The analysis incorporates five binary formation channels: \Belcpop, \ac{PopIII}, \ac{CHE}, \ac{GC}, and \ac{NSC}.}

The mixed model results demonstrate that isolated formation channels dominate \ac{BBH} production, accounting for approximately 90\% of astrophysical merger events. This finding is consistent with a previous study \cite{Zevin2021}, which also demonstrated that isolated dominance can break down when natal BH spins are large. In our analysis, we assume natal spins for isolated \ac{BBH}s are less than 0.2, making the observed \Belcpop dominance consistent with theoretical expectations.

\refine{In our mixed model analysis, the \ac{CHE} channel contributes only a small fraction to the detected population, despite its characteristic mass scale around $\sim 30\,M_\odot$ like the \ac{PopIII} channnel. This is mainly because the \ac{CHE} channnel predicts a broad positive $\chi_{\rm eff}$ distribution with a peak around $\chi_{\rm eff}\sim 0.5$, whereas the current GWTC-3 BBH sample does not favor such high effective spins. Therefore, the relative detectable fractions are governed by the combined effects of the mass distribution, spin distribution, and detector selection effects.}
In addition, we find that the \ac{PopIII} channel contributes more significantly than the dynamical formation channels (\ac{GC} and \ac{NSC}). However, our results show a stronger preference for the \Belcpop channel compared to \cite{Zevin2021}. This discrepancy may arise from differences in the underlying population synthesis models employed in each study.
This highlights a fundamental challenge in population inference studies: results can vary substantially depending on the choice of population synthesis framework. The sensitivity to model assumptions underscores the importance of continued refinement of theoretical predictions and the need for systematic comparison across different modeling approaches.

Our likelihood ratio analysis demonstrates that incorporating the \ac{PopIII} channel increases the model likelihood by a factor of $\sim$10. According to Jeffreys' scale for Bayes factors, this constitutes strong evidence, though it falls short of being decisive. However, this likelihood improvement must be interpreted cautiously due to the nested nature of our model comparison framework.
The observed enhancement can partially reflect the inherent advantage conferred by additional parameter flexibility rather than necessarily indicating superior theoretical validity. The inclusion of an additional formation channel naturally increases the model's capacity to accommodate the observed data through increased degrees of freedom. Consequently, the likelihood difference alone cannot definitively establish whether the \ac{PopIII} channel provides a more accurate or physically motivated explanation of the GWTC-3 observations.

\begin{figure}[htb!]
\begin{center}
\includegraphics[width=10cm]{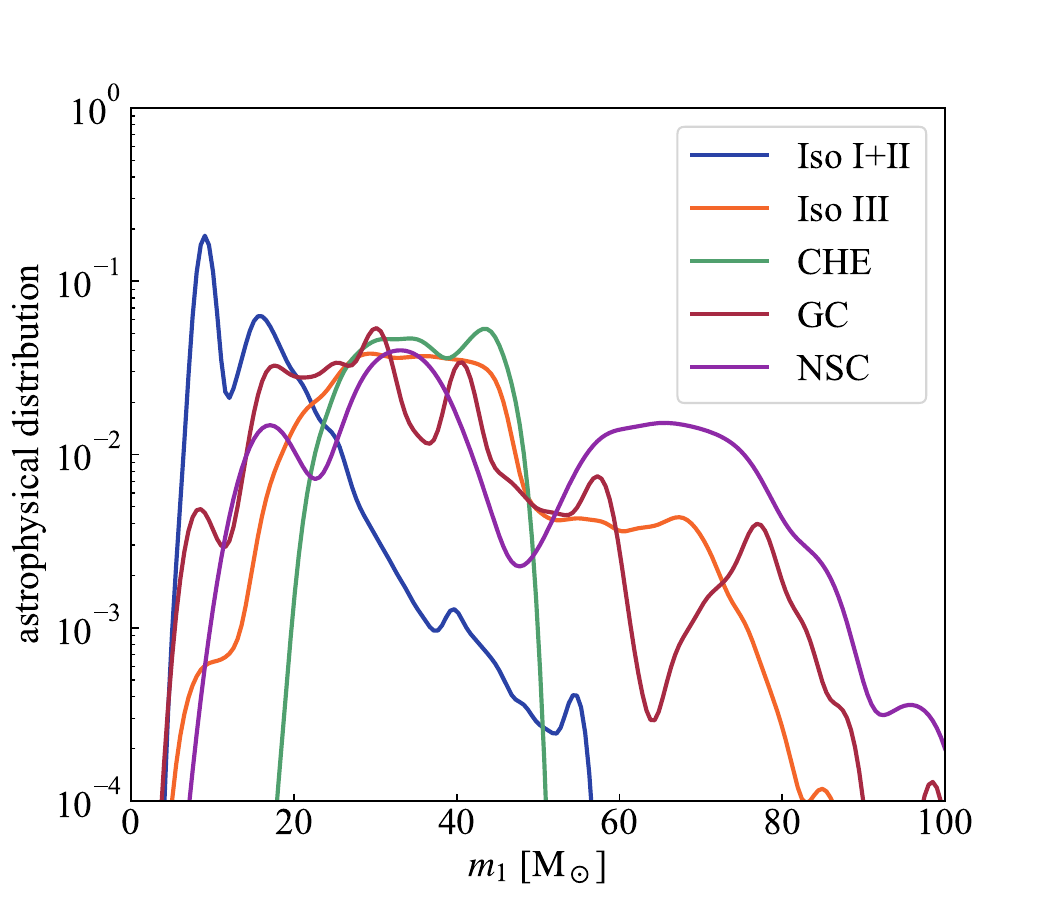}
\caption{The primary mass distribution of \ac{BBH} in the formation channels that are taken into account in this analysis.}
\label{fig: Individual_m1}
\end{center}
\end{figure}

\begin{figure}[htb!]
\begin{center}
\includegraphics[width=10cm]{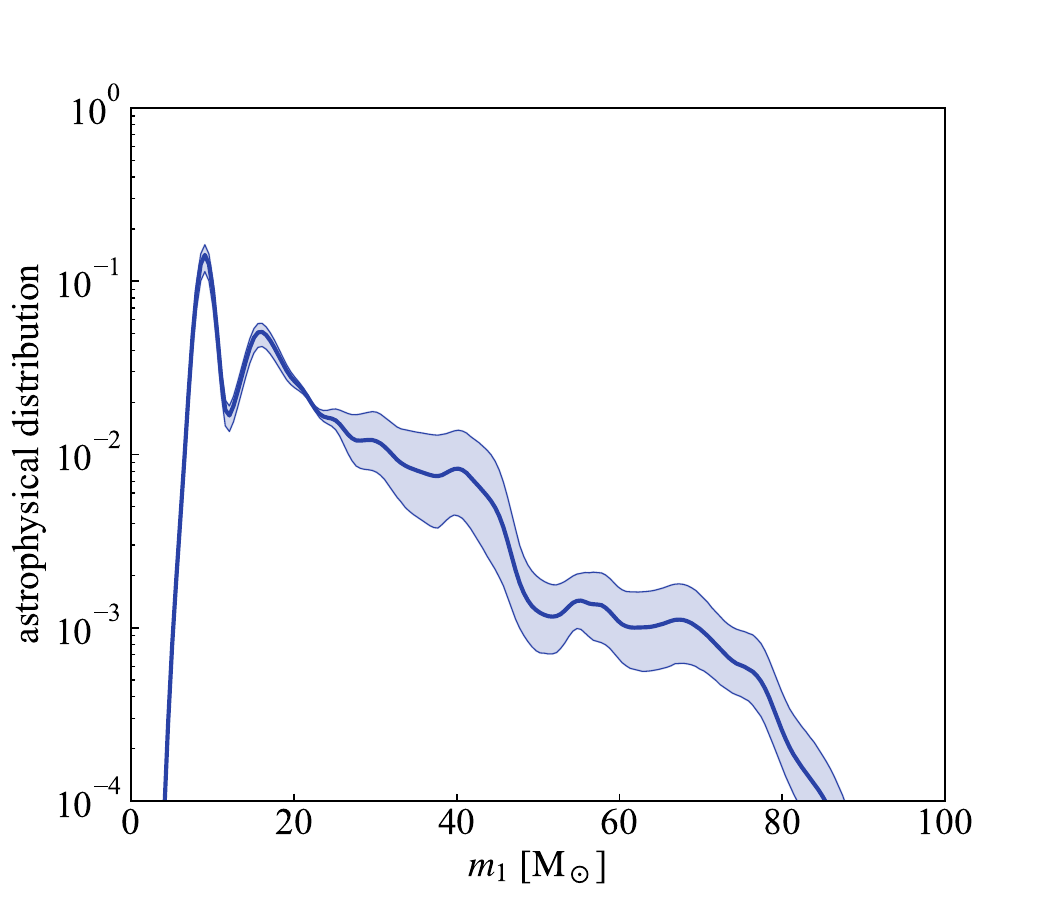}
\caption{The posterior predictive distribution for the primary mass distribution.}
\label{fig: Recovered_m1}
\end{center}
\end{figure}

Using the inferred relative fraction parameters, one can obtain astrophysical predictive distributions within this framework for physical quantities.

The mass distribution of formation channels that are used in this analysis is depicted in \figref{fig: Individual_m1}. \Belcpop channel mainly contributes to the BHs with mass $\leq20\ \Mo$. The contribution from \ac{CHE} models is concentrated at $m_1\approx30\ \Mo$. \ac{PopIII}, \ac{GC}, and \ac{NSC} channels contribute to the mass distribution for a wide mass range. \figref{fig: Recovered_m1} shows the recovered distribution with 90\% credibility from the synthesis of the formation channels. The $m_1$ distribution is highly peaked at $m_1\approx10\ \Mo$, from \Belcpop channel. Distribution of $m_1\gtrsim20\ \Mo$ is more uncertain compared to the less massive region, reflecting the uncertainty of the compositions of \ac{PopIII}, \ac{GC}, and \ac{NSC} channels.

\begin{figure}[htb!]
\begin{center}
\includegraphics[width=10cm]{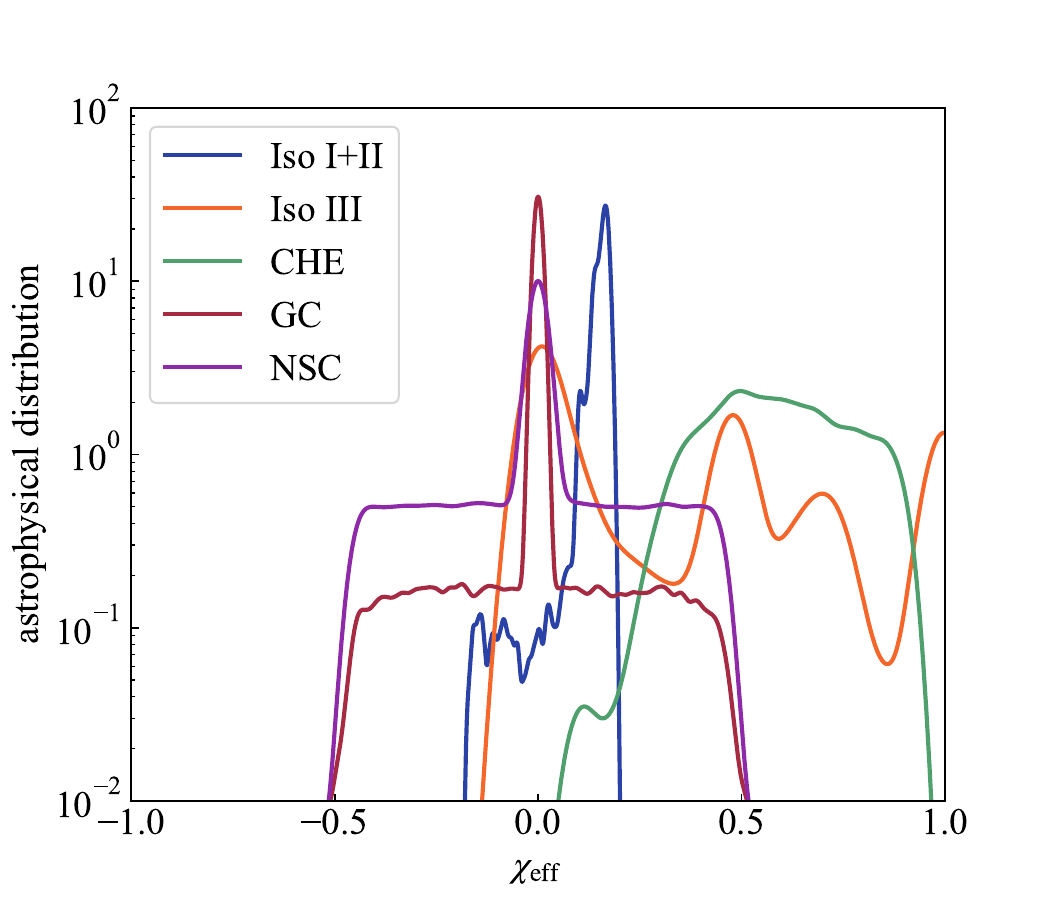}
\caption{Same as \figref{fig: Individual_m1}, but for $\Xeff$.}
\label{fig: Individual_Xeff}
\end{center}
\end{figure}

\begin{figure}[htb!]
\begin{center}
\includegraphics[width=10cm]{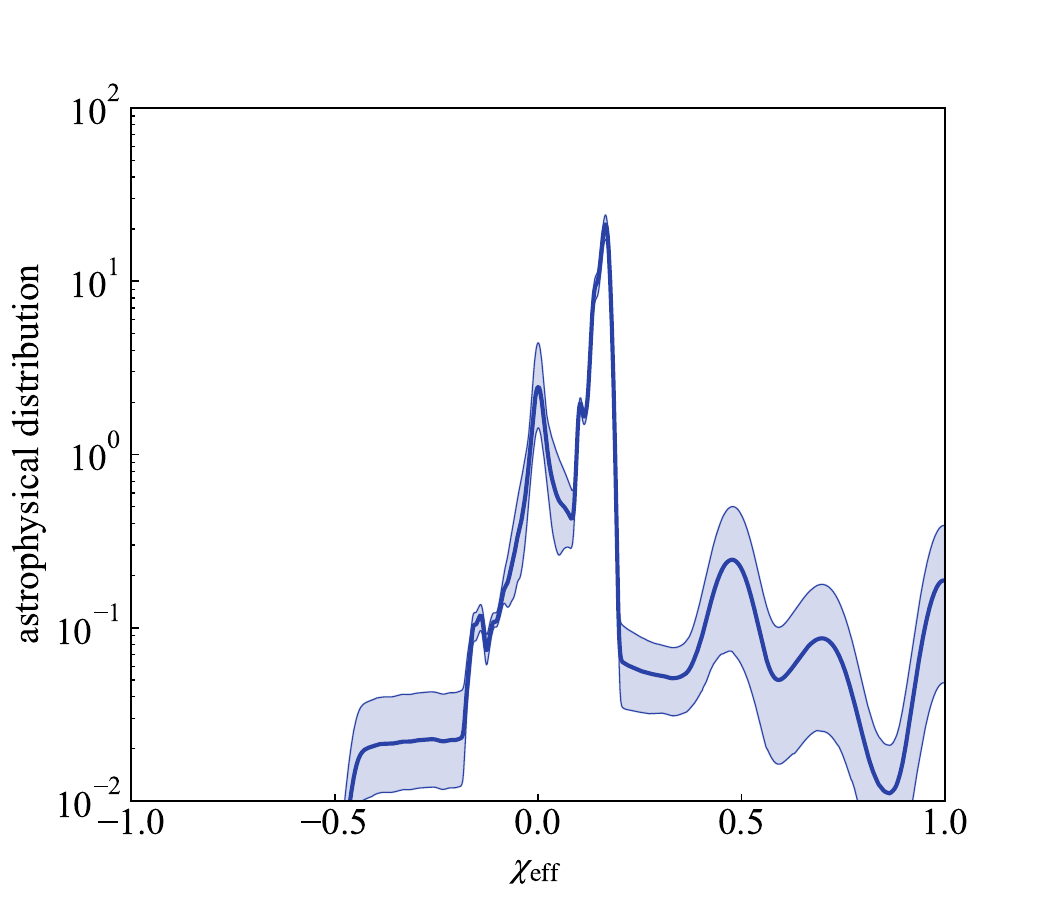}
\caption{Same as \figref{fig: Recovered_m1}, but for $\Xeff$.}
\label{fig: Recovered_Xeff}
\end{center}
\end{figure}

\figref{fig: Individual_Xeff} illustrates the $\Xeff$ distribution of formation channels. Though the treatment of the spin in population synthesis of isolated evolution has uncertainty, they tend to have positive $\Xeff$. This positive alignment is expected because the stellar spins of the progenitor stars are generally aligned with the orbital angular momentum in isolated binaries. The $\Xeff$ distribution for the CHE channel has a peak around $\Xeff\approx0.5$, and it is widely distributed with positive $\Xeff$ spanning from $\Xeff\approx0$ to $\Xeff\approx1$. This broad but positive distribution reflects the efficient angular momentum transfer in \ac{CHE} channel.

The $\Xeff$ distribution of dynamical evolutions can be decomposed as a peak at $\Xeff=0$ and a flat distribution around $\abs{\Xeff}\approx0.5$. The former can be explained by the spin orientation being randomized through many-body interactions in dense stellar environments, leading to no preferred spin alignment. The latter is the contribution from the hierarchical mergers, as the massive BH from a BH merger has a spin magnitude around $\abs{\chi}\approx 0.6$ due to the angular momentum conservation during the merger process. The \ac{NSC} channel tends to have more hierarchical mergers compared to the \ac{GC} channel, primarily because the deeper gravitational potential and higher escape velocity in \ac{NSC} allow them to retain more merger products.

\figref{fig: Recovered_Xeff} shows the $\Xeff$ distribution obtained from the population inference. The uncertainty of spin distribution at $\Xeff>0.2$ can be mainly attributed to the uncertainty of the fraction of the \ac{PopIII} channel, while no event to date has such a high $\Xeff$ parameter with high credibility. Thus, further subdivision of the isolated evolution channel would allow for more precise estimates. This approach could provide more detailed physical insight. 

Compared to the latest population inference from LVK collaboration \cite{GWTC-4:pop}, the predicted distribution for the low-mass end of $m_1$ exhibits the structure around $\sim 20\Mo$ noted in GWTC-4 in a more pronounced form, whilst a dip around $\sim15 \Mo$ has newly emerged. Within our framework, these characteristics are largely attributed to the \Belcpop model.
Stars with low metallicity ($\sim0.1Z_{\odot}$) tend to produce black holes with masses around $\sim20M_{\odot}$ \cite{2016A&A...594A..97B}, which gives rise to a local enhancement in the black-hole mass distribution.
For the high-mass end, the predictive distribution from our framework drops earlier than the LVK's one, but a relatively small number of such events makes the uncertainty of this region bigger. Further observations in this region are awaited.
%将来のイベントでより詳細な形が決まるだろう、という記述をする。
For the $\Xeff$ distribution, the discrepancy between this framework and GWTC-4 is much clearer. The distribution for $\Xeff$ in GWTC-4 analysis is fitted with a (skewed-)Gaussian distribution, while the predictive distribution from this framework shows multiple features because of the contributions of each subpopulation. If one considers the integrated subpopulations to be the true picture, the shape of the effective spin distribution appears highly jagged whch may have been smoothed out by the fit by a Gaussian-like distribution. However, further accumulation of events is awaited to clarify this.

\refine{We note that our results are subject to several sources of uncertainty.
For \ac{PopIII} stars, the assumed initial mass function (IMF) can affect the BH mass distribution, particularly the slope of the high-mass tail, although the characteristic mass scale
around $\sim 20$--$30\Mo$ is expected to be relatively insensitive to the IMF \citep{K2014,kinugawa2021}. For \Belcpop binaries, uncertainties in the treatment of the common-envelope phase can significantly impact the formation efficiency and orbital properties of merging systems \citep{Belczynski2002,B2008,Olejak:2021fti,Zevin2021}.
In addition, for dynamically formed binaries such as those in globular clusters, the assumed initial spins of BHs can affect the predicted spin properties of the population \citep{Zevin2021}.
These uncertainties may quantitatively affect our results, and a more systematic exploration is left for future work.}
In this study, we examine five formation pathways, but note that these selections are not determinative. Various other intriguing formation pathways exist in reality. One notable example of a formation channel not included in our analysis is the active galactic nucleus (AGN) channel. This channel considers the possibility that binary black holes (BBHs) form and merge within the dense gas disks surrounding supermassive black holes. 
We did not include the AGN channel in our mixture model since publicly available population synthesis data for this channel are not currently accessible. Nevertheless, several studies have suggested that the AGN channel could make a non-negligible contribution to the observed \ac{BBH} population, particularly for systems with high masses or moderate eccentricities \cite{Bartos2017, Yang2019}.  Due to the migration and interaction processes in AGN disks, \ac{BBH}s formed in such environments can exhibit both aligned and misaligned spin configurations.  This would lead to an intermediate distribution of the effective spin parameter, $\chi_{\mathrm{eff}}$, which differs from that predicted by isolated binary evolution or dense stellar dynamics.
Recent theoretical work by Tagawa et al.~\cite{Tagawa2020} demonstrates that repeated mergers of stellar-mass black holes can occur efficiently in AGN disks, leading to the formation of massive black holes through hierarchical mergers. 
Such processes could naturally produce black holes in the pair-instability mass gap and may also result in observable signatures in spin and mass distributions. 
Furthermore, recent hierarchical Bayesian analyses indicate that the AGN channel might explain some of the features seen in the GWTC-3 catalog, including possible hints of hierarchical mergers or spin misalignment \cite{Gayathri2023}. 
Although our current framework cannot directly constrain the AGN contribution, the inclusion of this channel in future studies—once detailed population models become available—will be important for a more complete understanding of \ac{BBH} formation channels.
Another dynamical formation channel not included in our current mixture model is the young dense cluster (YDC) channel.
YDCs are stellar systems with high densities and a rich population of massive stars during their early evolutionary phases.
Frequent dynamical interactions in such environments—particularly binary-single and binary-binary encounters—facilitate the efficient formation and hardening of binary black holes (BBHs), some of which may merge within the cluster's lifetime \cite{DiCarlo2020}.
A distinctive feature of YDC-origin \ac{BBH}s lies in their spin distribution.
Black holes that form directly from massive stellar collapse in YDCs are generally expected to have low natal spin magnitudes, assuming efficient angular momentum transport in progenitor stars.
However, YDCs are also favorable environments for hierarchical mergers: a black hole formed in an earlier merger may participate in subsequent mergers due to the high interaction rates.
Such hierarchical mergers typically produce remnant black holes with moderate to high spin ($a \sim 0.6$–$0.8$), depending on the mass ratio and spin alignment of the progenitor \ac{BBH}s \cite{Gerosa2017}.
The mass distribution of merging \ac{BBH}s in young dense cluster models appears to be an approximately log-flat distribution, i.e., $p(M) \propto M^{-1}$ from 5 $M_{\odot}$ to 50 $M_{\odot}$ \cite{DiCarlo2020}. 
Therefore, YDCs can host a population of massive black holes with higher spin than what is predicted by isolated binary evolution, while still allowing a broad distribution in the effective inspiral spin parameter $\chi_{\mathrm{eff}}$ due to random spin orientations acquired through dynamical processes.

Although we do not include the YDC channel in our current analysis due to the lack of publicly available population synthesis datasets specifically targeting this environment, the potential contribution of YDC-origin \ac{BBH}s—especially in the form of massive, high-spin systems—makes this channel a strong candidate for inclusion in future work.

Finally, since Pop~III stars are the first stars in the universe, they formed much earlier than Pop~I/II stars. Previous studies show that the peak of the Pop~III \ac{BBH} formation is located at $z \sim 10
$ \cite{Tanikawa2021,K2020chirp}.
Therefore, next-generation \ac{GW} detectors will reveal the nature of the \ac{BBH} population (such as the redshift-spin correlation) more clearly. The prediction of binary parameter distribution under the mixed model will be informative for the upcoming next-generation \ac{GW} detectors era.

\section{Summary}
\label{Sec.Sum}
There are many theoretical pathways proposed to explain the formation of \ac{BBH}s. In this paper, we take five formation channels: \Belcpop, \ac{PopIII}, \ac{CHE}, \ac{GC}, and \ac{NSC} channels.
From our mixing scheme revealed that \Belcpop channel dominates the origin of \ac{BBH}s, but it also suggested that \ac{PopIII} channel contributes approximately 10\% to \ac{BBH} mergers across the entire universe. From the perspective of observed catalogues, contributions from diverse formation channels, including channels beyond those mentioned above, were suggested. These findings underscore the importance of simultaneously considering the contributions of various formation processes. It should be noted, however, that the inferred branching fractions $\{f_j\}$ 
and the resulting posterior predictive distributions of the source parameters, 
such as the mass distribution, are inherently conditioned on the choice of 
formation channel models considered in this work. Each model encodes a set of 
astrophysical assumptions and idealizations, and the results may shift 
quantitatively as these prescriptions are refined or as additional formation 
channels are incorporated.
Future improvements in both theoretical models and observational constraints will be essential to refine our understanding of \ac{BBH} formation pathways. Further observations and the advent of next-generation detectors will lead to the accumulation of events with increasingly precise analysis results in the future. These events will enable the imposition of stricter constraints on the contribution rates from each channel. Furthermore, incorporating additional formation channels and reducing model uncertainties through multi-messenger observations and refined stellar evolution physics will enable us to construct a more complete picture of the astrophysical \ac{BBH} formation. Ultimately, disentangling the origins of \ac{BBH}s will not only illuminate the diverse evolutionary paths of massive stars and compact objects but also provide crucial insights into the star formation history and chemical evolution of the universe across cosmic time.

\section*{Acknowledgement}
M. I. was supported by Forefront Physics and Mathematics Program to Drive Transformation (FoPM), a World-leading Innovative Graduate Study (WINGS) Program, the University of Tokyo.
This research has made use of data obtained from the Gravitational Wave Open Science Center (gwosc.org), a service of LIGO Laboratory, the LIGO Scientific Collaboration, the Virgo Collaboration, and KAGRA. LIGO Laboratory and Advanced LIGO are funded by the United States National Science Foundation (NSF) as well as the Science and Technology Facilities Council (STFC) of the United Kingdom, the Max-Planck-Society (MPS), and the State of Niedersachsen/Germany for support of the construction of Advanced LIGO and construction and operation of the GEO600 detector. Additional support for Advanced LIGO was provided by the Australian Research Council. Virgo is funded, through the European Gravitational Observatory (EGO), by the French Centre National de Recherche Scientifique (CNRS), the Italian Istituto Nazionale di Fisica Nucleare (INFN) and the Dutch Nikhef, with contributions by institutions from Belgium, Germany, Greece, Hungary, Ireland, Japan, Monaco, Poland, Portugal, Spain. KAGRA is supported by Ministry of Education, Culture, Sports, Science and Technology (MEXT), Japan Society for the Promotion of Science (JSPS) in Japan; National Research Foundation (NRF) and Ministry of Science and ICT (MSIT) in Korea; Academia Sinica (AS) and National Science and Technology Council (NSTC) in Taiwan.
This work was supported by JSPS Grant-inAid for Scientific Research on Innovative Areas 2905: JP17H06358, JP17H06361 and JP17H06364, JSPS Core-to-Core Program A. Advanced Research Networks, JSPS Grant-in-Aid for Transformative Research Areas (A) 20A203: JP20H05854, the joint research program of the Institute for Cosmic Ray Research, University of Tokyo.

T. K. acknowledges support from JSPS KAKENHI Grant Numbers JP21K13915 and JP22K03630 and the financial support from the Science Moves Award.

\appendix
\section{Hierarchical Bayesian Analysis}
\label{Sec:HBDerivation}
Here, we describe a summary of the hierarchical Bayesian analysis method. In a Bayesian manner, given a set of data $\qty{x}$ from observation, one can calculate the posterior distribution for population parameter $\Lambda$, $p(\Lambda\mid\qty{x})$, as
\begin{equation}
p(\Lambda\mid\qty{x}) \propto \pi(\Lambda)p(\qty{x}\mid\Lambda),
\end{equation}
where $\pi(\Lambda)$ is the prior distribution for the population parameter $\Lambda$. If we assume that all events are independent of each other, this probability $p(\Lambda\mid\qty{x})$ can be expressed as
\begin{equation}
    p(\Lambda\mid\qty{x}) = \pi(\Lambda)\prod_{i=1}^{N_\mathrm{obs}}p(x_i\mid\Lambda),
\end{equation}
where $x_i$ denotes the observation data of $i$-th event. For each \ac{GW} event, we estimate the source parameter, such as masses, and we denote this by $\theta.$ The probability of detecting the \ac{GW} event $x$ from population parameter $\Lambda$ can be described by marginalizing over $\theta$:
\begin{equation}
\label{eq:naive_p}
    p(x\mid\Lambda)\propto \int \dd{\theta}p(x\mid\theta)p(\theta\mid\Lambda).
\end{equation}
Since we have the detection, $x$ must be 'detectable.' In other words, $x$ must pass the detection criteria. Therefore, the normalization for equation (\ref{eq:naive_p}) must be
\begin{equation}
\int \dd{x}p(x\mid\Lambda) = 1,
\end{equation}
where the integral domain is the entire $x$ satisfying detection criteria. This requires a parameter so-called 'detection efficiency' $\alpha(\Lambda)$ as a normalization factor:
\begin{equation}
\alpha(\Lambda) = \int\dd{\theta}p_\mathrm{det}(\theta)p(\theta\mid\Lambda),
\end{equation}
where $p_\mathrm{det}(\theta)$ is the probability that a gravitational wave from a binary system with binary parameter $\theta$ is observed by the detectors. With the normalization factor, the equation to assess $p(x\mid\Lambda)$ becomes
\begin{equation}
    p(x\mid\Lambda)= \dfrac{1}{\alpha(\Lambda)}\int \dd{\theta}p(x\mid\theta)p(\theta\mid\Lambda).
\end{equation}

For every event in the GWTC-3, LVK collaboration estimates the source parameter on it, applying Bayesian analysis \cite{GWTC-3}. Hence, $p(x\mid\theta)$ can be converted to $p(\theta\mid x)p(x)/\pi(\theta)$ by applying Bayes' theorem:
\begin{equation}
    p(x\mid\Lambda)= \dfrac{1}{\alpha(\Lambda)}\int \dd{\theta}\dfrac{p(x)p(\theta\mid x)}{\pi(\theta)}p(\theta\mid\Lambda).
\end{equation}

Finally, we have
\begin{equation}
    p(\Lambda\mid\qty{x})= \pi(\Lambda)\prod_{i=1}^{N_\mathrm{det}}\qty[\dfrac{1}{\alpha(\Lambda)}\int \dd{\theta}\dfrac{p(\theta\mid x_i)}{\pi(\theta)}p(\theta\mid\Lambda)].
\end{equation}

\bibliography{apssamp}%

@PREAMBLE{
 "\providecommand{\noopsort}[1]{}" 
 # "\providecommand{\singleletter}[1]{#1}%" 
}

@article{Bartos2017,
  author       = {Bartos, I. and Kocsis, B. and Haiman, Z. and Márka, S.},
  title        = {Merging Black Holes in Galactic Nuclei: Implications for Advanced LIGO Detections},
  journal      = {Astrophysical Journal},
  year         = {2017},
  volume       = {835},
  number       = {2},
  pages        = {165},
  doi          = {10.3847/1538-4357/835/2/165},
  archivePrefix= {arXiv},
  eprint       = {1602.03831},
  primaryClass = {astro-ph.HE}
}

@ARTICLE{Belczynski2002,
       author = {{Belczynski}, Krzysztof and {Kalogera}, Vassiliki and {Bulik}, Tomasz},
        title = "{A Comprehensive Study of Binary Compact Objects as Gravitational Wave Sources: Evolutionary Channels, Rates, and Physical Properties}",
      journal = {\apj},
         year = 2002,
        month = jun,
       volume = {572},
       number = {1},
        pages = {407-431},
          doi = {10.1086/340304},
archivePrefix = {arXiv},
       eprint = {astro-ph/0111452},
 primaryClass = {astro-ph},
       adsurl = {https://ui.adsabs.harvard.edu/abs/2002ApJ...572..407B}
}

@article{Olejak:2021fti,
    author = "Olejak, Aleksandra and Belczynski, Krzysztof and Ivanova, Natalia",
    title = "{Impact of common envelope development criteria on the formation of LIGO/Virgo sources}",
    eprint = "2102.05649",
    archivePrefix = "arXiv",
    primaryClass = "astro-ph.HE",
    doi = "10.1051/0004-6361/202140520",
    journal = "Astron. Astrophys.",
    volume = "651",
    pages = "A100",
    year = "2021"
}

@article{Gerosa2017,
  author = {Gerosa, Davide and Berti, Emanuele},
  title = {Are merging black holes born from stellar collapse or previous mergers?},
  journal = {Phys. Rev. D},
  volume = {95},
  pages = {124046},
  year = {2017},
  doi = {10.1103/PhysRevD.95.124046},
  eprint = {1703.06223},
  archivePrefix = {arXiv}
}

@ARTICLE{1944ApJ...100..137B,
       author = {{Baade}, W.},
        title = "{The Resolution of Messier 32, NGC 205, and the Central Region of the Andromeda Nebula.}",
      journal = {\apj},
         year = 1944,
        month = sep,
       volume = {100},
        pages = {137},
          doi = {10.1086/144650},
       adsurl = {https://ui.adsabs.harvard.edu/abs/1944ApJ...100..137B}
}

@ARTICLE{Rodriguez2016,
       author = {{Rodriguez}, Carl L. and {Chatterjee}, Sourav and {Rasio}, Frederic A.},
        title = "{Binary black hole mergers from globular clusters: Masses, merger rates, and the impact of stellar evolution}",
      journal = {\prd},
         year = 2016,
        month = apr,
       volume = {93},
       number = {8},
          eid = {084029},
        pages = {084029},
          doi = {10.1103/PhysRevD.93.084029},
archivePrefix = {arXiv},
       eprint = {1602.02444},
 primaryClass = {astro-ph.HE},
       adsurl = {https://ui.adsabs.harvard.edu/abs/2016PhRvD..93h4029R}
}

@ARTICLE{Banerjee2017,
       author = {{Banerjee}, Sambaran},
        title = "{Stellar-mass black holes in young massive and open stellar clusters and their role in gravitational-wave generation}",
      journal = {Mon. Not. Roy. Astron. Soc.},
         year = 2017,
        month = may,
       volume = {467},
       number = {1},
        pages = {524-539},
          doi = {10.1093/mnras/stw3392},
archivePrefix = {arXiv},
       eprint = {1611.09357},
 primaryClass = {astro-ph.HE},
       adsurl = {https://ui.adsabs.harvard.edu/abs/2017MNRAS.467..524B}
}

@ARTICLE{DiCarlo2019,
       author = {{Di Carlo}, Ugo N. and {Giacobbo}, Nicola and {Mapelli}, Michela and {Pasquato}, Mario and {Spera}, Mario and {Wang}, Long and {Haardt}, Francesco},
        title = "{Merging black holes in young star clusters}",
      journal = {Mon. Not. Roy. Astron. Soc.},
         year = 2019,
        month = aug,
       volume = {487},
       number = {2},
        pages = {2947-2960},
          doi = {10.1093/mnras/stz1453},
archivePrefix = {arXiv},
       eprint = {1901.00863},
 primaryClass = {astro-ph.HE},
       adsurl = {https://ui.adsabs.harvard.edu/abs/2019MNRAS.487.2947D}
}

@ARTICLE{Antonini2016,
       author = {{Antonini}, Fabio and {Rasio}, Frederic A.},
        title = "{Merging Black Hole Binaries in Galactic Nuclei: Implications for Advanced-LIGO Detections}",
      journal = {\apj},
         year = 2016,
        month = nov,
       volume = {831},
       number = {2},
          eid = {187},
        pages = {187},
          doi = {10.3847/0004-637X/831/2/187},
archivePrefix = {arXiv},
       eprint = {1606.04889},
 primaryClass = {astro-ph.HE},
       adsurl = {https://ui.adsabs.harvard.edu/abs/2016ApJ...831..187A}
}

@ARTICLE{Antonini2019,
       author = {{Antonini}, Fabio and {Gieles}, Mark and {Gualandris}, Alessia},
        title = "{Black hole growth through hierarchical black hole mergers in dense star clusters: implications for gravitational wave detections}",
      journal = {Mon. Not. Roy. Astron. Soc.},
         year = 2019,
        month = jul,
       volume = {486},
       number = {4},
        pages = {5008-5021},
          doi = {10.1093/mnras/stz1149},
archivePrefix = {arXiv},
       eprint = {1811.03640},
 primaryClass = {astro-ph.HE},
       adsurl = {https://ui.adsabs.harvard.edu/abs/2019MNRAS.486.5008A}
}

@ARTICLE{CHEmodel,
       author = {{du Buisson}, L. and {Marchant}, P. and {Podsiadlowski}, Ph and {Kobayashi}, C. and {Abdalla}, F.~B. and {Taylor}, P. and {Mandel}, I. and {de Mink}, S.~E. and {Moriya}, T.~J. and {Langer}, N.},
        title = "{Cosmic rates of black hole mergers and pair-instability supernovae from chemically homogeneous binary evolution}",
      journal = {Mon. Not. Roy. Astron. Soc.},
         year = 2020,
        month = dec,
       volume = {499},
       number = {4},
        pages = {5941-5959},
          doi = {10.1093/mnras/staa3225},
archivePrefix = {arXiv},
       eprint = {2002.11630},
 primaryClass = {astro-ph.HE},
       adsurl = {https://ui.adsabs.harvard.edu/abs/2020MNRAS.499.5941D}
}

@ARTICLE{Marchant2016,
       author = {{Marchant}, Pablo and {Langer}, Norbert and {Podsiadlowski}, Philipp and {Tauris}, Thomas M. and {Moriya}, Takashi J.},
        title = "{A new route towards merging massive black holes}",
      journal = {Astron. Astrophys.},
         year = 2016,
        month = apr,
       volume = {588},
          eid = {A50},
        pages = {A50},
          doi = {10.1051/0004-6361/201628133},
archivePrefix = {arXiv},
       eprint = {1601.03718},
 primaryClass = {astro-ph.SR},
       adsurl = {https://ui.adsabs.harvard.edu/abs/2016A&A...588A..50M}
}

@ARTICLE{K2016,
       author = {{Kinugawa}, Tomoya and {Miyamoto}, Akinobu and {Kanda}, Nobuyuki and {Nakamura}, Takashi},
        title = "{The detection rate of inspiral and quasi-normal modes of Population III binary black holes which can confirm or refute the general relativity in the strong gravity region}",
      journal = {Mon. Not. Roy. Astron. Soc.},
         year = 2016,
        month = feb,
       volume = {456},
       number = {1},
        pages = {1093-1114},
          doi = {10.1093/mnras/stv2624},
archivePrefix = {arXiv},
       eprint = {1505.06962},
 primaryClass = {astro-ph.SR},
       adsurl = {https://ui.adsabs.harvard.edu/abs/2016MNRAS.456.1093K}
}

@ARTICLE{2015ARA&A..53..631F,
       author = {{Frebel}, Anna and {Norris}, John E.},
        title = "{Near-Field Cosmology with Extremely Metal-Poor Stars}",
      journal = {Annu. Rev. Astron. Astrophys.},
         year = 2015,
        month = aug,
       volume = {53},
        pages = {631-688},
          doi = {10.1146/annurev-astro-082214-122423},
archivePrefix = {arXiv},
       eprint = {1501.06921},
 primaryClass = {astro-ph.SR},
       adsurl = {https://ui.adsabs.harvard.edu/abs/2015ARA&A..53..631F}
}

@ARTICLE{2016A&A...594A..97B,
       author = {{Belczynski}, K. and {Heger}, A. and {Gladysz}, W. and {Ruiter}, A.~J. and {Woosley}, S. and {Wiktorowicz}, G. and {Chen}, H.-Y. and {Bulik}, T. and {O'Shaughnessy}, R. and {Holz}, D.~E. and {Fryer}, C.~L. and {Berti}, E.},
        title = "{The effect of pair-instability mass loss on black-hole mergers}",
      journal = {Astron. Astrophys.},
         year = 2016,
        month = oct,
       volume = {594},
          eid = {A97},
        pages = {A97},
          doi = {10.1051/0004-6361/201628980},
archivePrefix = {arXiv},
       eprint = {1607.03116},
 primaryClass = {astro-ph.HE},
       adsurl = {https://ui.adsabs.harvard.edu/abs/2016A&A...594A..97B}
}

@article{Yang2019,
  author       = {Yang, Y. and Bartos, I. and Gayathri, V. and Ford, K. E. S. and Haiman, Z. and Klimenko, S. and Kocsis, B. and Márka, S. and Márka, Z. and McKernan, B. and O'Shaughnessy, R.},
  title        = {Hierarchical Black Hole Mergers in Active Galactic Nuclei},
  journal      = {Physical Review Letters},
  year         = {2019},
  volume       = {123},
  pages        = {181101},
  doi          = {10.1103/PhysRevLett.123.181101},
  archivePrefix= {arXiv},
  eprint       = {1906.09281},
  primaryClass = {astro-ph.HE}
}

@article{DiCarlo2020,
  author       = {Di Carlo, U. N. and Mapelli, M. and Giacobbo, N. and Spera, M. and Bouffanais, Y. and Rastello, S. and Santoliquido, F. and Pasquato, M. and Ballone, A. and Trani, A. A. and Torniamenti, S. and Haardt, F.},
  title        = {MOCCA-SURVEY Database I: Dynamical Formation of Merging Binary Black Holes in Young Star Clusters},
  journal      = {Mon. Not. Roy. Astron. Soc.},
  volume       = {498},
  number       = {1},
  pages        = {495--517},
  year         = {2020},
  doi          = {10.1093/mnras/staa2286},
  archivePrefix= {arXiv},
  eprint       = {2004.09525},
  primaryClass = {astro-ph.HE}
}

@article{Tagawa2020,
  author       = {Tagawa, H. and Umemura, M. and Gouda, M.},
  title        = {Multiple Mergers of Black Holes in Active Galactic Nucleus Disks: Effects of Disk Migration and Binary Formation},
  journal      = {Astrophysical Journal},
  volume       = {891},
  number       = {2},
  pages        = {119},
  year         = {2020},
  doi          = {10.3847/1538-4357/ab76c7},
  archivePrefix= {arXiv},
  eprint       = {1912.01699},
  primaryClass = {astro-ph.HE}
}

@article{Gayathri2023,
  author       = {Gayathri, V. and Wysocki, D. and Yang, Y. and Delfavero, V. and O'Shaughnessy, R. and Haiman, Z. and Tagawa, H. and Bartos, I.},
  title        = {Probing the Nature of the Gravitational Wave Source Population with GWTC-3},
  journal      = {Astrophysical Journal Letters},
  year         = {2023},
  volume       = {945},
  number       = {2},
  pages        = {L29},
  doi          = {10.3847/2041-8213/acbfb8},
  archivePrefix= {arXiv},
  eprint       = {2301.04187},
  primaryClass = {gr-qc}
}

@ARTICLE{K2020chirp,
       author = {{Kinugawa}, Tomoya and {Nakamura}, Takashi and {Nakano}, Hiroyuki},
        title = "{Chirp mass and spin of binary black holes from first star remnants}",
      journal = {Mon. Not. Roy. Astron. Soc.},
         year = 2020,
        month = nov,
       volume = {498},
       number = {3},
        pages = {3946-3963},
          doi = {10.1093/mnras/staa2511},
archivePrefix = {arXiv},
       eprint = {2005.09795},
 primaryClass = {astro-ph.HE},
       adsurl = {https://ui.adsabs.harvard.edu/abs/2020MNRAS.498.3946K}
}

@ARTICLE{Belczynskimodel,
       author = {{Belczynski}, K. and others},
        title = "{Evolutionary roads leading to low effective spins, high black hole masses, and O1/O2 rates for LIGO/Virgo binary black holes}",
      journal = {Astron. Astrophys.},
         year = 2020,
        month = apr,
       volume = {636},
          eid = {A104},
        pages = {A104},
          doi = {10.1051/0004-6361/201936528},
archivePrefix = {arXiv},
       eprint = {1706.07053},
 primaryClass = {astro-ph.HE},
       adsurl = {https://ui.adsabs.harvard.edu/abs/2020A&A...636A.104B}
}

@ARTICLE{GWTC-3,
       author = {{Abbott}, R. and others},
collaboration = {{LIGO Scientific Collaboration}, Virgo Collaboration, and {KAGRA Collaboration}},
        title = "{GWTC-3: Compact Binary Coalescences Observed by LIGO and Virgo during the Second Part of the Third Observing Run}",
      journal = {Physical Review X},
         year = 2023,
        month = oct,
       volume = {13},
       number = {4},
          eid = {041039},
        pages = {041039},
          doi = {10.1103/PhysRevX.13.041039},
archivePrefix = {arXiv},
       eprint = {2111.03606},
 primaryClass = {gr-qc},
       adsurl = {https://ui.adsabs.harvard.edu/abs/2023PhRvX..13d1039A}
}

@ARTICLE{GWOSC,
       author = {{Abbott}, R. and others},
       collaboration = {{LIGO Scientific Collaboration} and {Virgo Collaboration} and {KAGRA Collaboration}},
        title = "{Open Data from the Third Observing Run of LIGO, Virgo, KAGRA, and GEO}",
      journal = {The Astrophysical Journal Supplement Series},
         year = 2023,
        month = aug,
       volume = {267},
       number = {2},
          eid = {29},
        pages = {29},
          doi = {10.3847/1538-4365/acdc9f},
archivePrefix = {arXiv},
       eprint = {2302.03676},
 primaryClass = {gr-qc},
       adsurl = {https://ui.adsabs.harvard.edu/abs/2023ApJS..267...29A}
}

@ARTICLE{Tanikawa2021,
       author = {{Tanikawa}, Ataru and {Susa}, Hajime and {Yoshida}, Takashi and {Trani}, Alessandro A. and {Kinugawa}, Tomoya},
        title = "{Merger Rate Density of Population III Binary Black Holes Below, Above, and in the Pair-instability Mass Gap}",
      journal = {\apj},
         year = 2021,
        month = mar,
       volume = {910},
       number = {1},
          eid = {30},
        pages = {30},
          doi = {10.3847/1538-4357/abe40d},
archivePrefix = {arXiv},
       eprint = {2008.01890},
 primaryClass = {astro-ph.HE},
       adsurl = {https://ui.adsabs.harvard.edu/abs/2021ApJ...910...30T}
}

@ARTICLE{Zevin2021,
       author = {{Zevin}, Michael and {Bavera}, Simone S. and {Berry}, Christopher P.~L. and {Kalogera}, Vicky and {Fragos}, Tassos and {Marchant}, Pablo and {Rodriguez}, Carl L. and {Antonini}, Fabio and {Holz}, Daniel E. and {Pankow}, Chris},
        title = "{One Channel to Rule Them All? Constraining the Origins of Binary Black Holes Using Multiple Formation Pathways}",
      journal = {\apj},
         year = 2021,
        month = apr,
       volume = {910},
       number = {2},
          eid = {152},
        pages = {152},
          doi = {10.3847/1538-4357/abe40e},
archivePrefix = {arXiv},
       eprint = {2011.10057},
 primaryClass = {astro-ph.HE},
       adsurl = {https://ui.adsabs.harvard.edu/abs/2021ApJ...910..152Z}
}

@ARTICLE{HBcite,
       author = {{Mandel}, Ilya and {Farr}, Will M. and {Gair}, Jonathan R.},
        title = "{Extracting distribution parameters from multiple uncertain observations with selection biases}",
      journal = {Mon. Not. Roy. Astron. Soc.},
         year = 2019,
        month = jun,
       volume = {486},
       number = {1},
        pages = {1086-1093},
          doi = {10.1093/mnras/stz896},
archivePrefix = {arXiv},
       eprint = {1809.02063},
 primaryClass = {physics.data-an},
       adsurl = {https://ui.adsabs.harvard.edu/abs/2019MNRAS.486.1086M}
}

@INPROCEEDINGS{CE1,
       author = {{Paczynski}, B.},
        title = "{Common Envelope Binaries}",
    booktitle = {Structure and Evolution of Close Binary Systems},
         year = 1976,
       editor = {{Eggleton}, Peter and {Mitton}, Simon and {Whelan}, John},
       volume = {73},
        month = jan,
        pages = {75},
       adsurl = {https://ui.adsabs.harvard.edu/abs/1976IAUS...73...75P}
}

@ARTICLE{CE2,
       author = {{Bethe}, Hans A. and {Brown}, G.~E.},
        title = "{Evolution of Binary Compact Objects That Merge}",
      journal = {\apj},
         year = 1998,
        month = oct,
       volume = {506},
       number = {2},
        pages = {780-789},
          doi = {10.1086/306265},
archivePrefix = {arXiv},
       eprint = {astro-ph/9802084},
 primaryClass = {astro-ph},
       adsurl = {https://ui.adsabs.harvard.edu/abs/1998ApJ...506..780B}
}

@ARTICLE{CE3,
       author = {{R{\"o}pke}, Friedrich K. and {De Marco}, Orsola},
        title = "{Simulations of common-envelope evolution in binary stellar systems: physical models and numerical techniques}",
      journal = {Living Reviews in Computational Astrophysics},
         year = 2023,
        month = dec,
       volume = {9},
       number = {1},
          eid = {2},
        pages = {2},
          doi = {10.1007/s41115-023-00017-x},
archivePrefix = {arXiv},
       eprint = {2212.07308},
 primaryClass = {astro-ph.SR},
       adsurl = {https://ui.adsabs.harvard.edu/abs/2023LRCA....9....2R}
}

@ARTICLE{SMT1,
       author = {{van den Heuvel}, E.~P.~J. and {Portegies Zwart}, S.~F. and {de Mink}, S.~E.},
        title = "{Forming short-period Wolf-Rayet X-ray binaries and double black holes through stable mass transfer}",
      journal = {Mon. Not. Roy. Astron. Soc.},
         year = 2017,
        month = nov,
       volume = {471},
       number = {4},
        pages = {4256-4264},
          doi = {10.1093/mnras/stx1430},
archivePrefix = {arXiv},
       eprint = {1701.02355},
 primaryClass = {astro-ph.SR},
       adsurl = {https://ui.adsabs.harvard.edu/abs/2017MNRAS.471.4256V}
}

@ARTICLE{SMT2,
       author = {{Neijssel}, Coenraad J. and {Vigna-G{\'o}mez}, Alejandro and {Stevenson}, Simon and {Barrett}, Jim W. and {Gaebel}, Sebastian M. and {Broekgaarden}, Floor S. and {de Mink}, Selma E. and {Sz{\'e}csi}, Dorottya and {Vinciguerra}, Serena and {Mandel}, Ilya},
        title = "{The effect of the metallicity-specific star formation history on double compact object mergers}",
      journal = {Mon. Not. Roy. Astron. Soc.},
         year = 2019,
        month = dec,
       volume = {490},
       number = {3},
        pages = {3740-3759},
          doi = {10.1093/mnras/stz2840},
archivePrefix = {arXiv},
       eprint = {1906.08136},
 primaryClass = {astro-ph.SR},
       adsurl = {https://ui.adsabs.harvard.edu/abs/2019MNRAS.490.3740N}
}

@ARTICLE{CHE1,
       author = {{de Mink}, S.~E. and {Mandel}, I.},
        title = "{The chemically homogeneous evolutionary channel for binary black hole mergers: rates and properties of gravitational-wave events detectable by advanced LIGO}",
      journal = {Mon. Not. Roy. Astron. Soc.},
         year = 2016,
        month = aug,
       volume = {460},
       number = {4},
        pages = {3545-3553},
          doi = {10.1093/mnras/stw1219},
archivePrefix = {arXiv},
       eprint = {1603.02291},
 primaryClass = {astro-ph.HE},
       adsurl = {https://ui.adsabs.harvard.edu/abs/2016MNRAS.460.3545D}
}

@ARTICLE{GC1,
       author = {{Kulkarni}, S.~R. and {Hut}, Piet and {McMillan}, Steve},
        title = "{Stellar black holes in globular clusters}",
      journal = {\nat},
         year = 1993,
        month = jul,
       volume = {364},
       number = {6436},
        pages = {421-423},
          doi = {10.1038/364421a0},
       adsurl = {https://ui.adsabs.harvard.edu/abs/1993Natur.364..421K}
}

@ARTICLE{GC2,
       author = {{Rodriguez}, Carl L. and {Morscher}, Meagan and {Pattabiraman}, Bharath and {Chatterjee}, Sourav and {Haster}, Carl-Johan and {Rasio}, Frederic A.},
        title = "{Binary Black Hole Mergers from Globular Clusters: Implications for Advanced LIGO}",
      journal = {Physical Review Letters},
         year = 2015,
        month = jul,
       volume = {115},
       number = {5},
          eid = {051101},
        pages = {051101},
          doi = {10.1103/PhysRevLett.115.051101},
archivePrefix = {arXiv},
       eprint = {1505.00792},
 primaryClass = {astro-ph.HE},
       adsurl = {https://ui.adsabs.harvard.edu/abs/2015PhRvL.115e1101R}
}

@ARTICLE{AGN1,
       author = {{Bartos}, Imre and {Kocsis}, Bence and {Haiman}, Zolt{\'a}n and {M{\'a}rka}, Szabolcs},
        title = "{Rapid and Bright Stellar-mass Binary Black Hole Mergers in Active Galactic Nuclei}",
      journal = {\apj},
         year = 2017,
        month = feb,
       volume = {835},
       number = {2},
          eid = {165},
        pages = {165},
          doi = {10.3847/1538-4357/835/2/165},
archivePrefix = {arXiv},
       eprint = {1602.03831},
 primaryClass = {astro-ph.HE},
       adsurl = {https://ui.adsabs.harvard.edu/abs/2017ApJ...835..165B}
}

@ARTICLE{AGN2,
       author = {{Yang}, Y. and {Bartos}, I. and {Gayathri}, V. and {Ford}, K.~E.~S. and {Haiman}, Z. and {Klimenko}, S. and {Kocsis}, B. and {M{\'a}rka}, S. and {M{\'a}rka}, Z. and {McKernan}, B. and {O'Shaughnessy}, R.},
        title = "{Hierarchical Black Hole Mergers in Active Galactic Nuclei}",
      journal = {\prl},
         year = 2019,
        month = nov,
       volume = {123},
       number = {18},
          eid = {181101},
        pages = {181101},
          doi = {10.1103/PhysRevLett.123.181101},
archivePrefix = {arXiv},
       eprint = {1906.09281},
 primaryClass = {astro-ph.HE},
       adsurl = {https://ui.adsabs.harvard.edu/abs/2019PhRvL.123r1101Y}
}

@ARTICLE{YSC1,
       author = {{Di Carlo}, Ugo N. and {Mapelli}, Michela and {Giacobbo}, Nicola and {Spera}, Mario and {Bouffanais}, Yann and {Rastello}, Sara and {Santoliquido}, Filippo and {Pasquato}, Mario and {Ballone}, Alessandro and {Trani}, Alessandro A. and {Torniamenti}, Stefano and {Haardt}, Francesco},
        title = "{Binary black holes in young star clusters: the impact of metallicity}",
      journal = {Mon. Not. Roy. Astron. Soc.},
         year = 2020,
        month = oct,
       volume = {498},
       number = {1},
        pages = {495-506},
          doi = {10.1093/mnras/staa2286},
archivePrefix = {arXiv},
       eprint = {2004.09525},
 primaryClass = {astro-ph.HE},
       adsurl = {https://ui.adsabs.harvard.edu/abs/2020MNRAS.498..495D}
}

@ARTICLE{popIII-1,
       author = {{Abel}, Tom and {Bryan}, Greg L. and {Norman}, Michael L.},
        title = "{The Formation of the First Star in the Universe}",
      journal = {Science},
         year = 2002,
        month = jan,
       volume = {295},
       number = {5552},
        pages = {93-98},
          doi = {10.1126/science.295.5552.93},
archivePrefix = {arXiv},
       eprint = {astro-ph/0112088},
 primaryClass = {astro-ph},
       adsurl = {https://ui.adsabs.harvard.edu/abs/2002Sci...295...93A}
}

@ARTICLE{Triple1,
       author = {{Antonini}, Fabio and {Toonen}, Silvia and {Hamers}, Adrian S.},
        title = "{Binary Black Hole Mergers from Field Triples: Properties, Rates, and the Impact of Stellar Evolution}",
      journal = {\apj},
         year = 2017,
        month = jun,
       volume = {841},
       number = {2},
          eid = {77},
        pages = {77},
          doi = {10.3847/1538-4357/aa6f5e},
archivePrefix = {arXiv},
       eprint = {1703.06614},
 primaryClass = {astro-ph.GA},
       adsurl = {https://ui.adsabs.harvard.edu/abs/2017ApJ...841...77A}
}

@ARTICLE{Bromm2004,
       author = {{Bromm}, Volker and {Larson}, Richard B.},
        title = "{The First Stars}",
      journal = {Annu. Rev. Astron. Astrophys.},
         year = 2004,
        month = sep,
       volume = {42},
       number = {1},
        pages = {79-118},
          doi = {10.1146/annurev.astro.42.053102.134034},
archivePrefix = {arXiv},
       eprint = {astro-ph/0311019},
 primaryClass = {astro-ph},
       adsurl = {https://ui.adsabs.harvard.edu/abs/2004ARA&A..42...79B}
}

@ARTICLE{Marigo2003,
       author = {{Marigo}, P. and {Chiosi}, C. and {Kudritzki}, R. -P.},
        title = "{Zero-metallicity stars. II. Evolution of very massive objects with mass loss}",
      journal = {Astron. Astrophys.},
         year = 2003,
        month = feb,
       volume = {399},
        pages = {617-630},
          doi = {10.1051/0004-6361:20021756},
archivePrefix = {arXiv},
       eprint = {astro-ph/0212057},
 primaryClass = {astro-ph},
       adsurl = {https://ui.adsabs.harvard.edu/abs/2003A&A...399..617M}
}

@ARTICLE{PBH1,
       author = {{Bird}, Simeon and {Cholis}, Ilias and {Mu{\~n}oz}, Julian B. and {Ali-Ha{\"\i}moud}, Yacine and {Kamionkowski}, Marc and {Kovetz}, Ely D. and {Raccanelli}, Alvise and {Riess}, Adam G.},
        title = "{Did LIGO Detect Dark Matter?}",
      journal = {\prl},
         year = 2016,
        month = may,
       volume = {116},
       number = {20},
          eid = {201301},
        pages = {201301},
          doi = {10.1103/PhysRevLett.116.201301},
archivePrefix = {arXiv},
       eprint = {1603.00464},
 primaryClass = {astro-ph.CO},
       adsurl = {https://ui.adsabs.harvard.edu/abs/2016PhRvL.116t1301B}
}

@ARTICLE{2023popIII,
       author = {{Santoliquido}, Filippo and {Mapelli}, Michela and {Iorio}, Giuliano and {Costa}, Guglielmo and {Glover}, Simon C.~O. and {Hartwig}, Tilman and {Klessen}, Ralf S. and {Merli}, Lorenzo},
        title = "{Binary black hole mergers from population III stars: uncertainties from star formation and binary star properties}",
      journal = {Mon. Not. Roy. Astron. Soc.},
         year = 2023,
        month = sep,
       volume = {524},
       number = {1},
        pages = {307-324},
          doi = {10.1093/mnras/stad1860},
archivePrefix = {arXiv},
       eprint = {2303.15515},
 primaryClass = {astro-ph.GA},
       adsurl = {https://ui.adsabs.harvard.edu/abs/2023MNRAS.524..307S}
}

@ARTICLE{GW190521popIII,
       author = {{Liu}, Boyuan and {Bromm}, Volker},
        title = "{The Population III Origin of GW190521}",
      journal = {Astrophys. J. Lett.},
         year = 2020,
        month = nov,
       volume = {903},
       number = {2},
          eid = {L40},
        pages = {L40},
          doi = {10.3847/2041-8213/abc552},
archivePrefix = {arXiv},
       eprint = {2009.11447},
 primaryClass = {astro-ph.GA},
       adsurl = {https://ui.adsabs.harvard.edu/abs/2020ApJ...903L..40L}
}

@ARTICLE{PopIIIBelc,
       author = {{Belczynski}, Krzysztof and {Ryu}, Taeho and {Perna}, Rosalba and {Berti}, Emanuele and {Tanaka}, Takamitsu L. and {Bulik}, Tomasz},
        title = "{On the likelihood of detecting gravitational waves from Population III compact object binaries}",
      journal = {Mon. Not. Roy. Astron. Soc.},
         year = 2017,
        month = nov,
       volume = {471},
       number = {4},
        pages = {4702-4721},
          doi = {10.1093/mnras/stx1759},
archivePrefix = {arXiv},
       eprint = {1612.01524},
 primaryClass = {astro-ph.HE},
       adsurl = {https://ui.adsabs.harvard.edu/abs/2017MNRAS.471.4702B}
}

@ARTICLE{Hartwig,
       author = {{Hartwig}, Tilman and {Volonteri}, Marta and {Bromm}, Volker and {Klessen}, Ralf S. and {Barausse}, Enrico and {Magg}, Mattis and {Stacy}, Athena},
        title = "{Gravitational waves from the remnants of the first stars}",
      journal = {Mon. Not. Roy. Astron. Soc.},
         year = 2016,
        month = jul,
       volume = {460},
       number = {1},
        pages = {L74-L78},
          doi = {10.1093/mnrasl/slw074},
archivePrefix = {arXiv},
       eprint = {1603.05655},
 primaryClass = {astro-ph.GA},
       adsurl = {https://ui.adsabs.harvard.edu/abs/2016MNRAS.460L..74H}
}

@ARTICLE{ET,
       author = {{Punturo}, M. and others},
        title = "{The Einstein Telescope: a third-generation gravitational wave observatory}",
      journal = {Classical and Quantum Gravity},
         year = 2010,
        month = oct,
       volume = {27},
       number = {19},
          eid = {194002},
        pages = {194002},
          doi = {10.1088/0264-9381/27/19/194002},
       adsurl = {https://ui.adsabs.harvard.edu/abs/2010CQGra..27s4002P}
}

@ARTICLE{Mandel2022review,
       author = {{Mandel}, Ilya and {Broekgaarden}, Floor S.},
        title = "{Rates of compact object coalescences}",
      journal = {Living Reviews in Relativity},
         year = 2022,
        month = dec,
       volume = {25},
       number = {1},
          eid = {1},
        pages = {1},
          doi = {10.1007/s41114-021-00034-3},
archivePrefix = {arXiv},
       eprint = {2107.14239},
 primaryClass = {astro-ph.HE},
       adsurl = {https://ui.adsabs.harvard.edu/abs/2022LRR....25....1M}
}

@ARTICLE{Franciolini2022,
       author = {{Franciolini}, Gabriele and {Pani}, Paolo},
        title = "{Searching for mass-spin correlations in the population of gravitational-wave events: The GWTC-3 case study}",
      journal = {\prd},
         year = 2022,
        month = jun,
       volume = {105},
       number = {12},
          eid = {123024},
        pages = {123024},
          doi = {10.1103/PhysRevD.105.123024},
archivePrefix = {arXiv},
       eprint = {2201.13098},
 primaryClass = {astro-ph.HE},
       adsurl = {https://ui.adsabs.harvard.edu/abs/2022PhRvD.105l3024F}
}

@article{Farrell:2020zju,
    author = {Farrell, Eoin J. and Groh, Jose H. and Hirschi, Raphael and Murphy, Laura and Kaiser, Etienne and Ekstr{\"o}m, Sylvia and Georgy, Cyril and Meynet, Georges},
    title = "{Is GW190521 the merger of black holes from the first stellar generations?}",
    eprint = "2009.06585",
    archivePrefix = "arXiv",
    primaryClass = "astro-ph.SR",
    doi = "10.1093/mnrasl/slaa196",
    journal = "Mon. Not. Roy. Astron. Soc.",
    volume = "502",
    number = "1",
    pages = "L40--L44",
    year = "2021"
}

@ARTICLE{B2008,
       author = {{Belczynski}, Krzysztof and {Kalogera}, Vassiliki and {Rasio}, Frederic A. and {Taam}, Ronald E. and {Zezas}, Andreas and {Bulik}, Tomasz and {Maccarone}, Thomas J. and {Ivanova}, Natalia},
        title = "{Compact Object Modeling with the StarTrack Population Synthesis Code}",
      journal = {The Astrophysical Journal Supplement Series},
         year = 2008,
        month = jan,
       volume = {174},
       number = {1},
        pages = {223-260},
          doi = {10.1086/521026},
archivePrefix = {arXiv},
       eprint = {astro-ph/0511811},
 primaryClass = {astro-ph},
       adsurl = {https://ui.adsabs.harvard.edu/abs/2008ApJS..174..223B}
}

@ARTICLE{SFR,
       author = {{Madau}, Piero and {Fragos}, Tassos},
        title = "{Radiation Backgrounds at Cosmic Dawn: X-Rays from Compact Binaries}",
      journal = {\apj},
         year = 2017,
        month = may,
       volume = {840},
       number = {1},
          eid = {39},
        pages = {39},
          doi = {10.3847/1538-4357/aa6af9},
archivePrefix = {arXiv},
       eprint = {1606.07887},
 primaryClass = {astro-ph.GA},
       adsurl = {https://ui.adsabs.harvard.edu/abs/2017ApJ...840...39M}
}

@ARTICLE{Spruit,
       author = {{Spruit}, H.~C.},
        title = "{Dynamo action by differential rotation in a stably stratified stellar interior}",
      journal = {Astron. Astrophys.},
         year = 2002,
        month = jan,
       volume = {381},
        pages = {923-932},
          doi = {10.1051/0004-6361:20011465},
archivePrefix = {arXiv},
       eprint = {astro-ph/0108207},
 primaryClass = {astro-ph},
       adsurl = {https://ui.adsabs.harvard.edu/abs/2002A&A...381..923S}
}

@ARTICLE{K2014,
       author = {{Kinugawa}, Tomoya and {Inayoshi}, Kohei and {Hotokezaka}, Kenta and {Nakauchi}, Daisuke and {Nakamura}, Takashi},
        title = "{Possible indirect confirmation of the existence of Pop III massive stars by gravitational wave}",
      journal = {Mon. Not. Roy. Astron. Soc.},
         year = 2014,
        month = aug,
       volume = {442},
       number = {4},
        pages = {2963-2992},
          doi = {10.1093/mnras/stu1022},
archivePrefix = {arXiv},
       eprint = {1402.6672},
 primaryClass = {astro-ph.HE},
       adsurl = {https://ui.adsabs.harvard.edu/abs/2014MNRAS.442.2963K}
}

@ARTICLE{kinugawa2021,
       author = {{Kinugawa}, Tomoya and {Nakamura}, Takashi and {Nakano}, Hiroyuki},
        title = "{Gravitational waves from Population III binary black holes are consistent with LIGO/Virgo O3a data for the chirp mass larger than {\ensuremath{\sim}}20 M$_{{\ensuremath{\odot}}}$}",
      journal = {Mon. Not. Roy. Astron. Soc.},
         year = 2021,
        month = jun,
       volume = {504},
       number = {1},
        pages = {L28-L33},
          doi = {10.1093/mnrasl/slab032},
archivePrefix = {arXiv},
       eprint = {2103.00797},
 primaryClass = {astro-ph.HE},
       adsurl = {https://ui.adsabs.harvard.edu/abs/2021MNRAS.504L..28K}
}

@ARTICLE{GWTC-1,
       author = {{Abbott}, B.~P. and others},
collaboration = {{LIGO Scientific Collaboration} and {Virgo Collaboration}},
        title = "{GWTC-1: A Gravitational-Wave Transient Catalog of Compact Binary Mergers Observed by LIGO and Virgo during the First and Second Observing Runs}",
      journal = {Physical Review X},
         year = 2019,
        month = jul,
       volume = {9},
       number = {3},
          eid = {031040},
        pages = {031040},
          doi = {10.1103/PhysRevX.9.031040},
archivePrefix = {arXiv},
       eprint = {1811.12907},
 primaryClass = {astro-ph.HE},
       adsurl = {https://ui.adsabs.harvard.edu/abs/2019PhRvX...9c1040A}
}

@ARTICLE{GWTC-2.1,
  title={GWTC-2.1: Deep extended catalog of compact binary coalescences observed by LIGO and Virgo during the first half of the third observing run},
  author={Abbott, R and Abbott, TD and Acernese, F and Ackley, K and Adams, C and Adhikari, N and Adhikari, RX and Adya, VB and Affeldt, C and Agarwal, D and others},
  journal = {arXiv e-prints},
  year = 2021,
  month = aug,
  eid = {arXiv:2108.01045},
  pages = {arXiv:2108.01045},
  doi = {10.48550/arXiv.2108.01045},
  archivePrefix = {arXiv},
  eprint = {2108.01045},
  primaryClass = {gr-qc},
  adsurl = {https://ui.adsabs.harvard.edu/abs/2021arXiv210801045T}
}

@ARTICLE{GW150914,
       author = {{Abbott}, B.~P. and others},
collaboration = {{LIGO Scientific Collaboration} and {Virgo Collaboration}},
        title = "{Observation of Gravitational Waves from a Binary Black Hole Merger}",
      journal = {\prl},
         year = 2016,
        month = feb,
       volume = {116},
       number = {6},
          eid = {061102},
        pages = {061102},
          doi = {10.1103/PhysRevLett.116.061102},
archivePrefix = {arXiv},
       eprint = {1602.03837},
 primaryClass = {gr-qc},
       adsurl = {https://ui.adsabs.harvard.edu/abs/2016PhRvL.116f1102A}
}

@ARTICLE{Sasaki2016,
       author = {{Sasaki}, Misao and {Suyama}, Teruaki and {Tanaka}, Takahiro and {Yokoyama}, Shuichiro},
        title = "{Primordial Black Hole Scenario for the Gravitational-Wave Event GW150914}",
      journal = {\prl},
         year = 2016,
        month = aug,
       volume = {117},
       number = {6},
          eid = {061101},
        pages = {061101},
          doi = {10.1103/PhysRevLett.117.061101},
archivePrefix = {arXiv},
       eprint = {1603.08338},
 primaryClass = {astro-ph.CO},
       adsurl = {https://ui.adsabs.harvard.edu/abs/2016PhRvL.117f1101S}
}

@ARTICLE{2017PhRvD..96l3523A,
       author = {{Ali-Ha{\"\i}moud}, Yacine and {Kovetz}, Ely D. and {Kamionkowski}, Marc},
        title = "{Merger rate of primordial black-hole binaries}",
      journal = {\prd},
         year = 2017,
        month = dec,
       volume = {96},
       number = {12},
          eid = {123523},
        pages = {123523},
          doi = {10.1103/PhysRevD.96.123523},
archivePrefix = {arXiv},
       eprint = {1709.06576},
 primaryClass = {astro-ph.CO},
       adsurl = {https://ui.adsabs.harvard.edu/abs/2017PhRvD..96l3523A}
}

@ARTICLE{Inayoshi2017,
       author = {{Inayoshi}, Kohei and {Hirai}, Ryosuke and {Kinugawa}, Tomoya and {Hotokezaka}, Kenta},
        title = "{Formation pathway of Population III coalescing binary black holes through stable mass transfer}",
      journal = {Mon. Not. Roy. Astron. Soc.},
         year = 2017,
        month = jul,
       volume = {468},
       number = {4},
        pages = {5020-5032},
          doi = {10.1093/mnras/stx757},
archivePrefix = {arXiv},
       eprint = {1701.04823},
 primaryClass = {astro-ph.HE},
       adsurl = {https://ui.adsabs.harvard.edu/abs/2017MNRAS.468.5020I}
}

@BOOK{Silverman,
       author = {{Silverman}, B.~W.},
        title = "{Density estimation for statistics and data analysis}",
         year = 1986,
       adsurl = {https://ui.adsabs.harvard.edu/abs/1986desd.book.....S}
}

@ARTICLE{Stone2017,
       author = {{Stone}, Nicholas C. and {Metzger}, Brian D. and {Haiman}, Zolt{\'a}n},
        title = "{Assisted inspirals of stellar mass black holes embedded in AGN discs: solving the `final au problem'}",
      journal = {Mon. Not. Roy. Astron. Soc.},
         year = 2017,
        month = jan,
       volume = {464},
       number = {1},
        pages = {946-954},
          doi = {10.1093/mnras/stw2260},
archivePrefix = {arXiv},
       eprint = {1602.04226},
 primaryClass = {astro-ph.GA},
       adsurl = {https://ui.adsabs.harvard.edu/abs/2017MNRAS.464..946S}
}

@ARTICLE{Silsbee2017,
       author = {{Silsbee}, Kedron and {Tremaine}, Scott},
        title = "{Lidov-Kozai Cycles with Gravitational Radiation: Merging Black Holes in Isolated Triple Systems}",
      journal = {\apj},
         year = 2017,
        month = feb,
       volume = {836},
       number = {1},
          eid = {39},
        pages = {39},
          doi = {10.3847/1538-4357/aa5729},
archivePrefix = {arXiv},
       eprint = {1608.07642},
 primaryClass = {astro-ph.HE},
       adsurl = {https://ui.adsabs.harvard.edu/abs/2017ApJ...836...39S}
}

@ARTICLE{HBcite2,
       author = {{Thrane}, Eric and {Talbot}, Colm},
        title = "{An introduction to Bayesian inference in gravitational-wave astronomy: Parameter estimation, model selection, and hierarchical models}",
      journal = {
    Publications of the Astronomical Society of Australia},
         year = 2019,
        month = mar,
       volume = {36},
          eid = {e010},
        pages = {e010},
          doi = {10.1017/pasa.2019.2},
archivePrefix = {arXiv},
       eprint = {1809.02293},
 primaryClass = {astro-ph.IM},
       adsurl = {https://ui.adsabs.harvard.edu/abs/2019PASA...36...10T}
}

@ARTICLE{Tanikawathreshold,
       author = {{Tanikawa}, Ataru and {Yoshida}, Takashi and {Kinugawa}, Tomoya and {Takahashi}, Koh and {Umeda}, Hideyuki},
        title = "{Fitting formulae for evolution tracks of massive stars under extreme metal-poor environments for population synthesis calculations and star cluster simulations}",
      journal = {Mon. Not. Roy. Astron. Soc.},
         year = 2020,
        month = jul,
       volume = {495},
       number = {4},
        pages = {4170-4191},
          doi = {10.1093/mnras/staa1417},
archivePrefix = {arXiv},
       eprint = {1906.06641},
 primaryClass = {astro-ph.HE},
       adsurl = {https://ui.adsabs.harvard.edu/abs/2020MNRAS.495.4170T}
}

@ARTICLE{GWTC-3:pop,
       author = {{Abbott}, R. and others},
collaboration = {{LIGO Scientific Collaboration} and {VIRGO Collaboration} and {KAGRA Collaboration}},
        title = "{Population of Merging Compact Binaries Inferred Using Gravitational Waves through GWTC-3}",
      journal = {Physical Review X},
         year = 2023,
        month = jan,
       volume = {13},
       number = {1},
          eid = {011048},
        pages = {011048},
          doi = {10.1103/PhysRevX.13.011048},
archivePrefix = {arXiv},
       eprint = {2111.03634},
 primaryClass = {astro-ph.HE},
       adsurl = {https://ui.adsabs.harvard.edu/abs/2023PhRvX..13a1048A}
}

@ARTICLE{Precision_Requirement,
       author = {{Essick}, Reed and {Farr}, Will},
        title = "{Precision Requirements for Monte Carlo Sums within Hierarchical Bayesian Inference}",
      journal = {arXiv e-prints},
         year = 2022,
        month = apr,
          eid = {arXiv:2204.00461},
        pages = {arXiv:2204.00461},
          doi = {10.48550/arXiv.2204.00461},
archivePrefix = {arXiv},
       eprint = {2204.00461},
 primaryClass = {astro-ph.IM},
       adsurl = {https://ui.adsabs.harvard.edu/abs/2022arXiv220400461E}
}

@ARTICLE{GWTC-4,
       author = {{The LIGO Scientific Collaboration} and {the Virgo Collaboration} and {the KAGRA Collaboration}},
        title = "{GWTC-4.0: Updating the Gravitational-Wave Transient Catalog with Observations from the First Part of the Fourth LIGO-Virgo-KAGRA Observing Run}",
      journal = {arXiv e-prints},
         year = 2025,
        month = aug,
          eid = {arXiv:2508.18082},
        pages = {arXiv:2508.18082},
          doi = {10.48550/arXiv.2508.18082},
archivePrefix = {arXiv},
       eprint = {2508.18082},
 primaryClass = {gr-qc},
       adsurl = {https://ui.adsabs.harvard.edu/abs/2025arXiv250818082T}
}

@ARTICLE{GWTC-4:pop,
       author = {{The LIGO Scientific Collaboration} and {the Virgo Collaboration} and {the KAGRA Collaboration}},
        title = "{GWTC-4.0: Population Properties of Merging Compact Binaries}",
      journal = {arXiv e-prints},
         year = 2025,
        month = aug,
          eid = {arXiv:2508.18083},
        pages = {arXiv:2508.18083},
          doi = {10.48550/arXiv.2508.18083},
archivePrefix = {arXiv},
       eprint = {2508.18083},
 primaryClass = {astro-ph.HE},
       adsurl = {https://ui.adsabs.harvard.edu/abs/2025arXiv250818083T}
}

@ARTICLE{NonparaCallister,
       author = {{Callister}, Thomas A. and {Farr}, Will M.},
        title = "{Parameter-Free Tour of the Binary Black Hole Population}",
      journal = {Physical Review X},
         year = 2024,
        month = apr,
       volume = {14},
       number = {2},
          eid = {021005},
        pages = {021005},
          doi = {10.1103/PhysRevX.14.021005},
archivePrefix = {arXiv},
       eprint = {2302.07289},
 primaryClass = {astro-ph.HE},
       adsurl = {https://ui.adsabs.harvard.edu/abs/2024PhRvX..14b1005C}
}

@ARTICLE{qXeffCorrelation,
       author = {{Callister}, Thomas A. and {Haster}, Carl-Johan and {Ng}, Ken K.~Y. and {Vitale}, Salvatore and {Farr}, Will M.},
        title = "{Who Ordered That? Unequal-mass Binary Black Hole Mergers Have Larger Effective Spins}",
      journal = {Astrophysical Journal Letters},
         year = 2021,
        month = nov,
       volume = {922},
       number = {1},
          eid = {L5},
        pages = {L5},
          doi = {10.3847/2041-8213/ac2ccc},
archivePrefix = {arXiv},
       eprint = {2106.00521},
 primaryClass = {astro-ph.HE},
       adsurl = {https://ui.adsabs.harvard.edu/abs/2021ApJ...922L...5C}
}

@INCOLLECTION{MapelliReivew,
       author = {{Mapelli}, Michela},
        title = "{Formation Channels of Single and Binary Stellar-Mass Black Holes}",
    booktitle = {Handbook of Gravitational Wave Astronomy},
         year = 2021,
       editor = {{Bambi}, Cosimo and {Katsanevas}, Stavros and {Kokkotas}, Konstantinos D.},
          eid = {16},
        pages = {16},
          doi = {10.1007/978-981-15-4702-7_16-1},
       adsurl = {https://ui.adsabs.harvard.edu/abs/2021hgwa.bookE..16M}
}

@ARTICLE{MandelFarmerReview,
       author = {{Mandel}, Ilya and {Farmer}, Alison},
        title = "{Merging stellar-mass binary black holes}",
      journal = {Phys. Rep.},
         year = 2022,
        month = apr,
       volume = {955},
        pages = {1-24},
          doi = {10.1016/j.physrep.2022.01.003},
archivePrefix = {arXiv},
       eprint = {1806.05820},
 primaryClass = {astro-ph.HE},
       adsurl = {https://ui.adsabs.harvard.edu/abs/2022PhR...955....1M}
}

\end{document}